\PassOptionsToClass{
  nofootinbib,
  citeautoscript,
  superscriptaddress,
  nobalancelastpage,
  floatfix
}{revtex4-1}

\documentclass[reprint]{revtex4-1}

\usepackage{graphicx}
\usepackage{siunitx}
\DeclareSIUnit\cycle{cyc}
\usepackage{fancyhdr}
\usepackage{mathtools}
\DeclarePairedDelimiter\abs{\lvert}{\rvert}
\usepackage[colorlinks=true,
    citecolor=blue,
    linkcolor=black,
    urlcolor=blue,
    pagebackref=false]{hyperref}
\usepackage[usenames,dvipsnames]{xcolor}
\definecolor{paleaqua}{rgb}{0.9, 0.95, 1}
\usepackage[version=4]{mhchem}
\usepackage{times}
\usepackage{bm}

\renewcommand{\d}[2]
  {\frac{d #1}{d #2}}

\newcommand{\pd}[2]
  {\frac{\partial #1}{\partial #2}} 
 
\newcommand{\pdc}[3]
  {\left( 
     \frac{\partial #1}{\partial #2}
  \right)_{#3}}

\renewcommand{\Re}{\operatorname{Re}}
\renewcommand{\Im}{\operatorname{Im}}

\newcommand{\latin}[1]{#1}

\newcommand{\st}[1]{_\mathrm{#1}}
\newcommand{\stt}[1]{_{\mbox{\tiny{#1}}}}
\newcommand{\sRi}{_{\mbox{\tiny Ri}}}
\newcommand{\sRs}{_{\mbox{\tiny Rs}}}

\newcommand{\figref}[1]{Fig.~\ref{#1}}
\newcommand{\eqnref}[1]{Eq.~(\ref{#1})}

\newcommand{\eqnsref}[1]{Eqs.~(\ref{#1})}

\begin{document}

  \pagestyle{fancy}

    \lhead{\footnotesize
      \textsf{DWYER, HARRELL, and MAROHN}}
    \rhead{\footnotesize \textsf{\today}}
    \lfoot{}
    \cfoot{\thepage}
    \rfoot{}
        
  \title{Lagrangian and impedance spectroscopy treatments of electric force microscopy}

  \author{Ryan P. Dwyer}
  \affiliation{Department of Chemistry and Chemical Biology, Cornell University, Ithaca NY 14853}
  \affiliation{Department of Chemistry and Biochemistry, University of Mount Union, Alliance OH 44601}

  \author{Lee E. Harrell}
  \affiliation{Department of Physics and Nuclear Engineering, U.S. Military Academy, West Point, NY 10996}

  \author{John A. Marohn}
  \affiliation{Department of Chemistry and Chemical Biology, Cornell University, Ithaca NY 14853}
  
\begin{abstract}
Scanning probe microscopy is often extended beyond simple topographic imaging to study electrical forces and sample properties, with the most widely used experiment being frequency-modulated Kelvin probe force microscopy.
The equations commonly used to interpret this frequency-modulated experiment, however, rely on two hidden assumptions.
The first assumption is that the tip charge oscillates in phase with the cantilever motion to keep the tip voltage constant.
The second assumption is that any changes in the tip-sample interaction happen slowly.
Starting from an electro-mechanical model of the cantilever-sample interaction, we use Lagrangian mechanics to derive coupled equations of motion for the cantilever position and charge.
We solve these equations analytically using perturbation theory, and, for verification, numerically.
This general approach rigorously describes scanned probe experiments even in the case when the usual assumptions of fast tip charging and slowly changing samples properties are violated.
We develop a Magnus-expansion approximation to illustrate how abrupt changes in the tip-sample interaction cause abrupt changes in the cantilever amplitude and phase.
We show that feedback-free time-resolved electric force microscopy cannot uniquely determine sub-cycle photocapacitance dynamics.
We then use first-order perturbation theory to relate cantilever frequency shift and dissipation to the sample impedance even when the tip charge oscillates out of phase with the cantilever motion.
Analogous to the treatment of impedance spectroscopy in electrochemistry, we apply this approximation to determine the cantilever frequency shift and dissipation for an arbitrary sample impedance in both local dielectric spectroscopy and broadband local dielectric spectroscopy experiments.
The general approaches we develop provide a path forward for rigorously modeling the coupled motion of the cantilever position and charge in the wide range of electrical scanned probe microscopy experiments where the hidden assumptions of the conventional equations are violated or inapplicable.
\end{abstract}

\date{\today}

\maketitle
\thispagestyle{fancy}

\section{Introduction}
\label{Sec:Introduction} 
\sloppy

The invention of the atomic-force microscope \cite{Binnig1986mar} (AFM) led to an explosion of microcantilever-based electric force microscope (EFM) experiments\footnote{For lack of a better moniker, let us use the term \emph{electric force microscope} to describe scanned-probe microscope experiments in which a voltage is applied to a microcantilever with the goal of measuring the electrical properties of a sample.} capable of mapping the electrical properties of a thin-film sample \cite{Kalinin2005}.
In spite of this progress, a unified,  rigorous theory describing the electro-mechanical forces at play in such experiments is lacking.
Here we present a unified Lagrangian treatment of the coupled motion of the cantilever position, cantilever charge, and sample charges in an electric force microscope experiment.
This treatment describes a wide variety of transient and steady-state experiments and reveals the hidden assumptions underlying many of the equations widely used by practitioners of electric force microscopy.

To appreciate why such a new treatment is helpful, consider a non-contact scanned-probe microscope experiment in which an electrically conductive cantilever having a sharp tip is used to measure the electrical properties of a conducting or semi-conducting sample --- the so-called scanning Kelvin probe force microscope  (KPFM) experiment \cite{Martin1988mar,Nonnenmacher1991jun,Weaver1991may,Kikukawa1995jun,Burgi2002apr,Burgi2003nov,Silveira2007}.
The cantilever is brought near a sample surface and is driven into resonant oscillation by applying a mechanical force to the base of the cantilever.
A voltage, either static or oscillating, is applied to the cantilever, and the cantilever's position or frequency is measured.
The cantilever's position is shifted by an electrostatic force acting on the charged tip.
This force is usually stated as
\begin{equation}
F 
  = \frac{1}{2}
  \frac{\partial C}{\partial z}
  \left( V - \Phi \right)^2,
  \label{eq:F-SKPM-const-V}
\end{equation}
with $z$ the axis of cantilever motion, $C$ the tip-sample capacitance, $V$ the applied voltage, and $\Phi$ the sample's surface potential.
The associated electrostatic force gradient shifts the cantilever's resonance frequency.
This shift is usually stated as
\begin{equation}
\Delta f 
  = - \frac{f_0}{4 k_0} 
  \frac{\partial^2 C}{\partial z^2}
  \left( V - \Phi\right)^2,
  \label{eq:Df-SKPM-const-V}
\end{equation}
with $f_0$ the cantilever resonance frequency and $k_0$ the cantilever spring constant.

While universally used, Eqs.~(\ref{eq:F-SKPM-const-V}) and (\ref{eq:Df-SKPM-const-V}) make assumptions about charge motion in the sample that are seldom explicitly stated  or experimentally checked.
In the remaining paragraphs of this introduction we summarize prior theoretical and experimental work questioning the validity of Eqs.~(\ref{eq:F-SKPM-const-V}) and (\ref{eq:Df-SKPM-const-V}) and summarize the new equations resulting from our Lagrangian treatment of electric force microscopy.

Silveira, Dunlap, and coworkers took up the question of how to rigorously derive Eqs.~(\ref{eq:F-SKPM-const-V}) and (\ref{eq:Df-SKPM-const-V}) \cite{Silveira2007}.
As a concrete starting point for  subsequent discussion, let us briefly reproduce their analysis here.
In the idealized description of the electric force microscope experiment presented in Ref.~\citenum{Silveira2007}, the sample is grounded and the cantilever-sample system is modeled as a parallel-plate capacitor.
A charge $q$ is transferred from the sample to the tip as a result of the applied voltage $V$ and the difference in the electron chemical potential of the cantilever tip and the sample ($\mu_{\mathrm{t}}$ and $\mu_{\mathrm{s}}$, respectively).
The energy needed to charge the associated cantilever-sample capacitor is given by the Helmholtz free energy
\begin{equation}
A(q,T) 
  = \frac{q^2}{2 C} 
  + \frac{q}{e} \, \Delta \mu,
\end{equation}
with $T$ temperature, $C$ the tip-sample capacitance, $e$ the electron charge, and $\Delta \mu = \mu_{\mathrm{s}} - \mu_{\mathrm{t}}$.
The first term in this equation accounts for the energy stored in the capacitor's electric field while the second term accounts for the change in free energy associated with transferring electrons between two different materials.
The tip-sample force \emph{at constant charge} is
\begin{equation}
F_{q} 
  = - \left( 
    \frac{\partial A}{\partial z} 
    \right)_{q,T} 
  = \frac{1}{2} 
  \frac{1}{C^2} 
  \frac{\partial C}{\partial z} q^2.
  \label{eq:Fq-Dunlap}
\end{equation}
With $z$ defined such that $z$ increases as the tip moves away from the sample, the capacitance derivative $\partial C / \partial z$ is negative.
The cantilever therefore feels a negative, attractive force, as one would expect from Coulomb's law since the tip and sample are oppositely charged.
For a parallel-plate capacitor, the capacitance depends on plate separation $z$ as $C \sim 1/z$ and consequently $\partial C / \partial z \sim -1/z^2$.

Computing the tip-sample force in a constant-voltage experiment requires additional analysis.
The voltage is defined as the variable which is conjugate to the charge, 
\begin{equation}
V = \left( 
    \frac{\partial A}{\partial q}
  \right)_{z,T} 
  = \frac{q}{C} + \frac{\Delta \mu}{e}.
  \label{eq:V-conj}
\end{equation}
When the cantilever is set to vibrate, $C$ will become time dependent and the charge will redistribute between the plates.
If the charge-redistribution time constant is much faster than the cantilever period, $q(t) = C(t) \, (V - \Delta \mu / e)$, and the system will maintain the tip at constant voltage continuously. 
Assuming this is the case, the force may be computed from the grand-canonical free energy, obtained through a Legendre transformation: $\Omega(V,T,z) = A - q V$,
where in writing $\Omega$ we must eliminate $q$ as the dependent variable.
The term $- q V$ accounts for the work required to move the charge through
the battery that maintains the tip at constant potential.
The force experienced by the cantilever
held \emph{at constant voltage} is obtained by differentiating the resulting grand-canonical free energy,
\begin{equation}
F_{V} 
  = - \left( 
    \frac{\partial \Omega}{\partial z} 
    \right)_{V,T} 
  = \frac{1}{2} 
  \frac{\partial C}{\partial z} 
  \left( 
    V - \frac{\Delta \mu}{e}
  \right)^2.
  \label{eq:FV-Dunlap}
\end{equation}
The capacitance derivative is negative and the cantilever feels a negative, attractive force when held at constant voltage.
Equation~(\ref{eq:F-SKPM-const-V}) reduces to Eq.~(\ref{eq:FV-Dunlap}) in the limit where $\Delta \mu/e = \Phi$.
For the case of a more interesting sample, $\Phi$ contains contributions from the sample's local electrostatic potential as well as the difference $\Delta \mu/e$ in the chemical potential of the tip and the sample's metallic contact \cite{Silveira2007}.
Equation~(\ref{eq:Df-SKPM-const-V}) is obtained from Eq.~(\ref{eq:FV-Dunlap}) by expanding the force in a Taylor series about an equilibrium position, identifying the $z$-dependent force as a spring constant shift that modifies the cantilever's resonance frequency and neglecting any higher-order terms.

The Silveira-Dunlap analysis reveals that Eqs.~(\ref{eq:F-SKPM-const-V}) and (\ref{eq:Df-SKPM-const-V}) implicitly assume that charge redistributes instantaneously between the tip and the sample as the cantilever moves.
In other words, as the cantilever vibrates sinusoidally, it is assumed that the tip and sample charges oscillate perfectly in phase with the sinusoidal motion of the cantilever.
How valid is this assumption in practice?
Early in the development of the electric force microscope, Denk and Pohl argued that currents induced in the sample by the oscillating cantilever would lead to Joule dissipation of energy at a rate that depended on the sample's local conductivity \cite{Denk1991oct} (expressed in terms of the spreading resistance \cite{Denhoff2006may}).
The energy lost to this Joule heating was supplied by the cantilever, leading to a cantilever dissipation dependent on the electrical conductivity of the sample below the tip.
Stowe \emph{et al.}, motivated by this idea, used cantilever dissipation to image the concentration of dopants in silicon \cite{Stowe1999nov}.
The postulated Joule heating underlying both of these experiments implies the existence of sample charge oscillating out of phase with the sinusoidal motion of the cantilever, calling into question the general validity of Eqs.~(\ref{eq:F-SKPM-const-V}) and (\ref{eq:Df-SKPM-const-V}).

This out-of-phase component of the oscillating sample charge has since been exploited to create striking EFM images of individual quantum dots \cite{Woodside2002may,Zhu2005dec,Zhu2008aug,Bennett2010jan,Cockins2010may,Cockins2012feb,Roy-Gobeil2015apr}.
These experiments relied on the dots operating in the Coulomb-blockade limit such that scanning the tip's dc voltage or position resulted in a step change in the number of electrons $n$ residing on the quantum dot.
Adjusting the tip voltage or height to operate near a $n \rightarrow n \pm 1$ transition, individual electrons could be pushed on and off the quantum dot by applying a small modulation to the tip voltage or height.
Due to the finite rate at which electrons tunneled on and off the dot, the electrostatic force acting on the cantilever caused a frequency shift and dissipation.
Characteristic oscillations in frequency shift and dissipation were seen as the tip's dc voltage or position was scanned and individual electrons were forced on or off the dot.

Sample-induced dissipation effects have been detected in a number of other experiments on semiconducting samples.
A bias-dependent contact friction was observed over gallium arsenide and modeled as arising from interactions of the tip with trapped charge in the sample \cite{Qi2008may}.
A measurable increase in non-contact friction was observed when a polymer-fullerene solar-cell film was illuminated with light, inducing photochemical damage \cite{Cox2013nov}.
Dramatic, simultaneous changes in cantilever frequency and dissipation were observed over an illuminated lead-halide perovskite sample; these changes were used to follow the slow relaxation of the sample's photocapacitance in the dark in real time \cite{Tirmzi2017jan} and the activation energy of the underlying relaxation process was measured by repeating the experiment at various temperatures.

\begin{figure}
\includegraphics{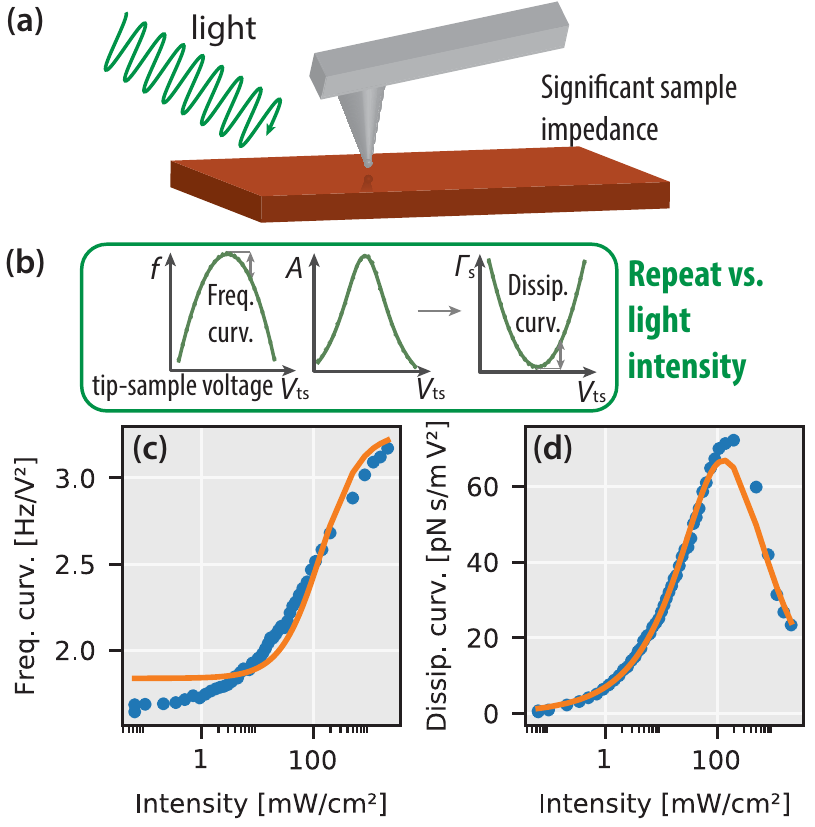}
\caption{Apparent violation of \eqnref{eq:Df-SKPM-const-V} in an illuminated thin-film semiconductor, \ce{CsPbBr3}.
(a) Experimental schematic.
(b) The cantilever frequency shift $\Delta f$ and amplitude $A$ are measured as a function of the tip-sample voltage $V\st{ts}$ and the sample-induced dissipation $\Gamma\st{s}$ is calculated from $A$.
The curvature of the (c) $\Delta f$-\latin{vs}-$V\st{ts}$ and (d) $\Gamma\st{s}$-\latin{vs}-$V\st{ts}$ parabolas versus illumination intensity.
The circles are measured data and the lines are a fits to the Lagrangian-impedance model discussed in the text.
The plots in (c,d) are adapted with permission from Ref.~\citenum{Tirmzi2017jan} (copyright 2017, American Chemical Society).
}
\label{fig:Tirmzi2017jan-expt}
\end{figure}

The illuminated-perovskite experiment is sketched in Fig.~\ref{fig:Tirmzi2017jan-expt}.
The sample is a thin-film semiconductor, \ce{CsPbBr3}, prepared on a conductive indium tin oxide substrate and illuminated from above with visible light.
The tip-sample capacitance derivative $C^{\prime\prime}$ and surface potential $\Phi$ are inferred, in the usual way, by measuring the cantilever frequency shift $\Delta f$ versus tip-sample voltage $V\st{ts}$.
According to \eqnref{eq:Df-SKPM-const-V}, the curvature of the $\Delta f$-\latin{vs}-$V\st{ts}$ parabola is $-f_0 C^{\prime\prime} / 4 k_0$, proportional to $C^{\prime\prime}$.
In a semiconductor sample like \ce{CsPbBr3} the free carrier density and therefore the capacitance should be proportional to the illumination intensity $I_{h\nu}$; we consequently expect to see a power-law dependence of $C^{\prime\prime}$ on  $I_{h\nu}$, which was not observed.
The cantilever dissipation $\Gamma\st{s}$ was also measured \latin{versus} $V\st{ts}$ and illumination intensity.
Here we likewise expect to see a power-law dependence of $\Gamma\st{s}$ on $I_{h\nu}$ with the dissipation increasing continuously with free-carrier density.
Instead, as $I_{h\nu}$ was increased linearly, the observed voltage-normalized dissipation increased, reached a maximum, and then decreased.

How can we explain this non-monotonic behavior?
In contrast with the quantum-dot experiments of Refs.~\citenum{Woodside2002may,Zhu2005dec,Zhu2008aug,Bennett2010jan,Cockins2010may,Cockins2012feb,Roy-Gobeil2015apr}, we cannot rely on Coulomb-blockade physics to describe the frequency-shift and dissipation effects seen in the semiconductor experiments of Refs.~\citenum{Qi2008may,Cox2013nov,Tirmzi2017jan}.
Moreover, we need to model the sample as a continuous film, ideally using a complex, frequency-dependent impedance.
Such an approach has been used to treat a number of related experiments.
In impedance microscopy measurements \cite{Shao2003mar,OHayre2004jun,OHayre2004sep,Lee2012jul} the tip is brought into contact with the sample and employed as the top capacitor plate in a conventional impedance spectroscopy measurement; modeling the signal in these experiments is straightforward because the cantilever is not moving.
Theoretical treatments of more sophisticated charged-cantilever measurements like local dielectric spectroscopy \cite{Crider2007jul,Crider2008jan}, broadband local dielectric spectroscopy \cite{Labardi2016may}, piezoresponse force microscopy \cite{Kalinin2004nov,Harnagea2003jul,Jesse2006jul,Eliseev2014jun}, and electrochemical strain microscopy \cite{Balke2010aug,Morozovska2010sep,NatalyChen2012aug} likewise treat the sample using a complex dielectric function, but fail to fully treat the coupled motion of sample charge and cantilever charge induced by the oscillation of the cantilever's  position and voltage.  

To understand the data of Fig.~\ref{fig:Tirmzi2017jan-expt}(c,d) we describe the sample using a complex impedance while employing a Lagrangian formalism to describe the coupled motion of the cantilever displacement, tip charge, and sample charge.
Applying this treatment to the Fig.~\ref{fig:Tirmzi2017jan-expt} experiment, below in Sec.~\ref{sec:Impedance_spectroscopy_EFM_theory} we obtain the frequency shift
\begin{equation}
\Delta f = -\frac{f_0}{4k_0} \Big( C''_q + \Delta C'' \Re\big(\hat{H}(\omega_0)\big) \Big ) V^2
  \label{Eq:KPFM-new-major-equation-I-preview}
\end{equation}
and sample-induced dissipation
\begin{equation}
\Gamma\st{s} = -\frac{1}{4 \pi f_0} \Delta C'' \, \Im\big(\hat{H}(\omega_0) \big) \, V^2,
  \label{Eq:KPFM-new-major-equation-II-preview}
\end{equation}
with $\Delta C'' = 2 (C^{\prime})^2/C$ and $C''_q = C'' - \Delta C''$ two distinct capacitance derivatives, and
\begin{equation}
\hat{H}(\omega) 
  = \frac{1}{1 + j \omega C Z(\omega)}
  \label{Eq:KPFM-new-major-equation-III-preview}
\end{equation}
a transfer function that depends on the tip capacitance and the complex sample impedance $Z(\omega)$.
In Ref.~\citenum{Tirmzi2017jan}, Tirmzi, Dwyer, and coworkers derived Eqs.~(\ref{Eq:KPFM-new-major-equation-I-preview}--\ref{Eq:KPFM-new-major-equation-III-preview}) by considering the components of the electrostatic force in-phase and out-of-phase with the oscillating cantilever.
Here we show these equations follow from a more general Lagrangian treatment which reveals the implicit assumptions undergirding Eqs.~(\ref{Eq:KPFM-new-major-equation-I-preview}--\ref{Eq:KPFM-new-major-equation-III-preview}).
These equations are one of the primary findings of this manuscript.
Equation~(\ref{Eq:KPFM-new-major-equation-I-preview}) should be used in place of \eqnref{eq:Df-SKPM-const-V} for semiconductors and other finite-impedance samples.
The physical insight we gain from these equations is that the frequency shift and dissipation probe the real and imaginary value, respectively, of the \eqnref{Eq:KPFM-new-major-equation-III-preview} transfer function at the cantilever oscillation frequency.

To explain the Fig.~\ref{fig:Tirmzi2017jan-expt}(c,d) data using Eqs.~(\ref{Eq:KPFM-new-major-equation-I-preview}) and (\ref{Eq:KPFM-new-major-equation-II-preview}) we model the sample as a capacitor $C\st{s}$ and light-dependent resistor $R\st{s}$ operating in parallel.
In this model, the transfer function in \eqnref{Eq:KPFM-new-major-equation-III-preview} has a roll-off frequency determined by the time constant $\tau = R\st{s}(C + C\st{s}) \approx R\st{s} C$.  
The non-monotonic behavior of $\Gamma\st{s}$ can be understood qualitatively as follows: in the Fig.~\ref{fig:Tirmzi2017jan-expt} experiment, $R\st{s}$ is large in the dark and small under illumination; the peak in $\Gamma\st{s}$ occurs at an illumination intensity where $2 \pi \tau$ matches the cantilever period.
The lines in Fig.~\ref{fig:Tirmzi2017jan-expt}(c,d) are a fit of the data to Eqs.~(\ref{Eq:KPFM-new-major-equation-I-preview}) and (\ref{Eq:KPFM-new-major-equation-II-preview}) assuming a sample time constant $\tau \propto I_{h\nu}^{-n}$ with $n = 0.6$, close to the value of $n = 0.5$ expected for photogenerated free carriers.
The joint fit nicely captures the nonlinear behavior of both the frequency and the dissipation versus illumination intensity.

\begin{figure}
\includegraphics{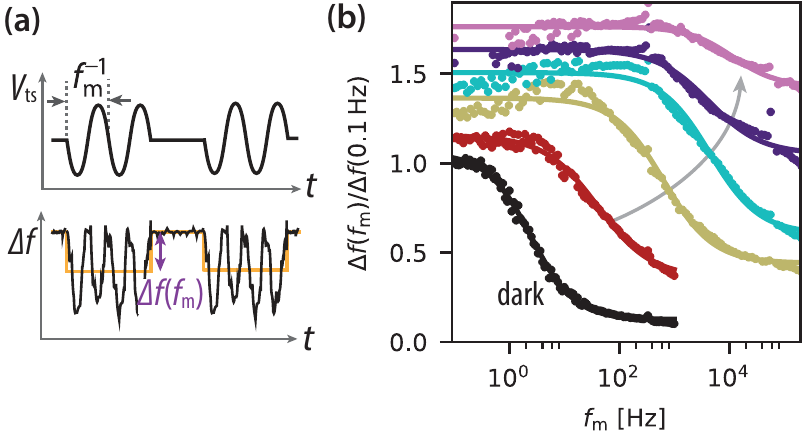}
\caption{Broadband local dielectric spectroscopy.
(a) Experimental schematic.
(b) Broadband local dielectric spectra collected at various light intensities over the semiconducting \ce{CsPbBr3} sample of Fig.~\ref{fig:Tirmzi2017jan-expt} \cite{Tirmzi2017jan}.
The plot in (b) is adapted with permission from Ref.~\citenum{Tirmzi2017jan} (copyright 2017, American Chemical Society).
}
\label{fig:BLDS-intro}
\end{figure}

The full frequency dependence of the \eqnref{Eq:KPFM-new-major-equation-III-preview} transfer function can be measured directly using a broadband local dielectric spectroscopy (BLDS) measurement (Fig.~\ref{fig:BLDS-intro}).
In one version of the experiment, the tip voltage is switched slowly on and off and, when on, is a sine wave of frequency $f\st{m}$.
The observed cantilever oscillation is sent to a frequency demodulator and the resulting time-dependent cantilever frequency shift sent to a lock-in amplifier with reference frequency set to the on/off modulation frequency.
The resulting signal, indicated as $\Delta f(f\st{m})$ in Fig.~\ref{fig:BLDS-intro}(a), changes when $f\st{m}$ is slowly varied. 
Using our Lagrangian-impedance formalism to calculate the measured frequency shift in such BLDS experiments, below in Sec.~\ref{sec:Impedance_spectroscopy_EFM_theory} we obtain
\begin{equation}
\Delta f (f\st{m}) = -\frac{f_0 V\st{m}^2}{16 k_0} \Big [
C''_q + \Delta C'' \Re \big( \bar{H} (\omega\st{m}, \omega_{0}) \big ) 
\Big ]
\abs{\hat{H}(\omega\st{m})}^2,
\label{eq:Deltaf-BLDS-intro}
\end{equation}
with $V\st{m}$ and $f\st{m}$ the amplitude and frequency of the applied oscillating voltage (assumed sinusoidal) and $\bar{H}$ the average value of the transfer function at frequencies $\omega\st{m} \pm \omega_{0}$.
In deriving \eqnref{eq:Deltaf-BLDS-intro} we assume for simplicity that a sinusoidal, not on/off, amplitude modulation is employed.
In Fig.~\ref{fig:BLDS-intro}(b) we show the BLDS frequency-shift spectrum measured at various light intensities over the semiconducting \ce{CsPbBr3} sample of Fig.~\ref{fig:Tirmzi2017jan-expt}.
The change in the spectrum's knee with increasing light intensity is in qualitative agreement with the light-dependent $R\st{s}$ used to explain the Fig.~\ref{fig:Tirmzi2017jan-expt} data, validating the use of a relatively simple $R C$ sample impedance model in explaining a wide range of experiments.

The treatment of transient effects in electrostatic force microscopy requires great care.
In time-domain EFM experiments the response of ions to a step-change in tip voltage is tracked in real time through a shift in cantilever frequency \cite{Bennewitz1997oct,Schirmeisen2004sep,Schirmeisen2007may,Taskiran2009jul,Schirmeisen2010,Mascaro2017jun}.
EFM has been used to follow the time evolution of photocapacitance in response to illumination \cite{Coffey2006sep,Cox2015aug,Giridharagopal2012jan,Karatay2016may}.
These experiments have pushed the limits of time resolution in EFM, with claimed time resolutions down to less than 1 percent of the cantilever period \cite{Karatay2016may}.
These EFM photocapacitance experiments stand in contrast to scanning probe microscopy-based variants of optical pump probe techniques, which exploit a nonlinearity
to infer ultrafast dynamics by measuring differences in a time-averaged quantity \latin{versus} a pulse time, delay, or frequency \cite{Hamers1990nov,Nunes1993nov}.
Recent experiments along these lines have measured the surface photovoltage \cite{Takihara2008jul,Shao2014oct,Schumacher2016apr,Schumacher2017jan} and charge moving through a transistor \cite{Murawski2015dec,Murawski2015oct} with ultrafast time resolution.
In contrast, the origin of sub-cycle time resolution in single-shot, transient EFM experiments is not clearly understood.

\begin{figure}
\includegraphics[width=3.32in]{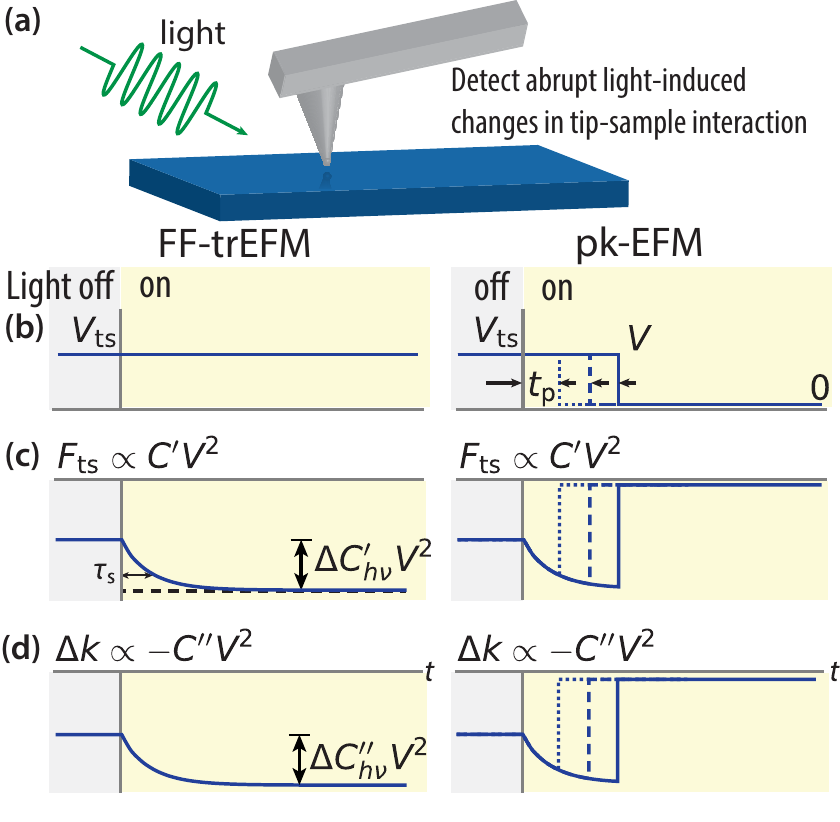}
\caption{FF-trEFM and pk-EFM timing diagrams.
(a) Experimental cartoon.
Both experiments start when the light is turned on.
(b) The tip-sample voltage during FF-tr-EFM is constant; during pk-EFM, the voltage is abruptly stepped to zero after a variable time $t\st{p}$.
In both FF-trEFM and pk-EFM, the tip-sample force $F\st{ts}$ (c) and the spring constant shift $\Delta k$ (d) change, possibly abruptly, after the start of the light pulse. 
The simplest model assumes the dynamics are single-exponential with a risetime $\tau\st{s}$.
In pk-EFM, the step change in voltage causes $F\st{ts}$ and $\Delta k$ to return to zero at delay time $t=t\st{p}$ .
}
\label{fig:trEFM-and-pkEFM-schematic}
\end{figure}

Two representative transient EFM measurements are shown in Fig.~\ref{fig:trEFM-and-pkEFM-schematic}.
The left side of Fig.~\ref{fig:trEFM-and-pkEFM-schematic} shows the feedback-free time-resolved electric force microscopy (FF-trEFM) experiment \cite{Karatay2016may} and the right side shows the phasekick electric force microscopy (pk-EFM) experiment \cite{Dwyer2017jun}.
The objective of both experiments is to observe the temporal dynamics of light-induced changes in a semiconductor sample's capacitance.
Applying light initiates a sample-related change in the tip-sample capacitance derivatives $C^{\prime}$ and $C^{\prime\prime}$ which for simplicity are sketched as a single-exponential with risetime $\tau\st{s}$.
In the presence of a finite tip voltage $V\st{ts}$, transients in $C^{\prime}$ and $C^{\prime\prime}$ induce the indicated transients in the tip-sample force $F\st{ts}$ and force gradient $\Delta k$.

\begin{figure*}
\includegraphics{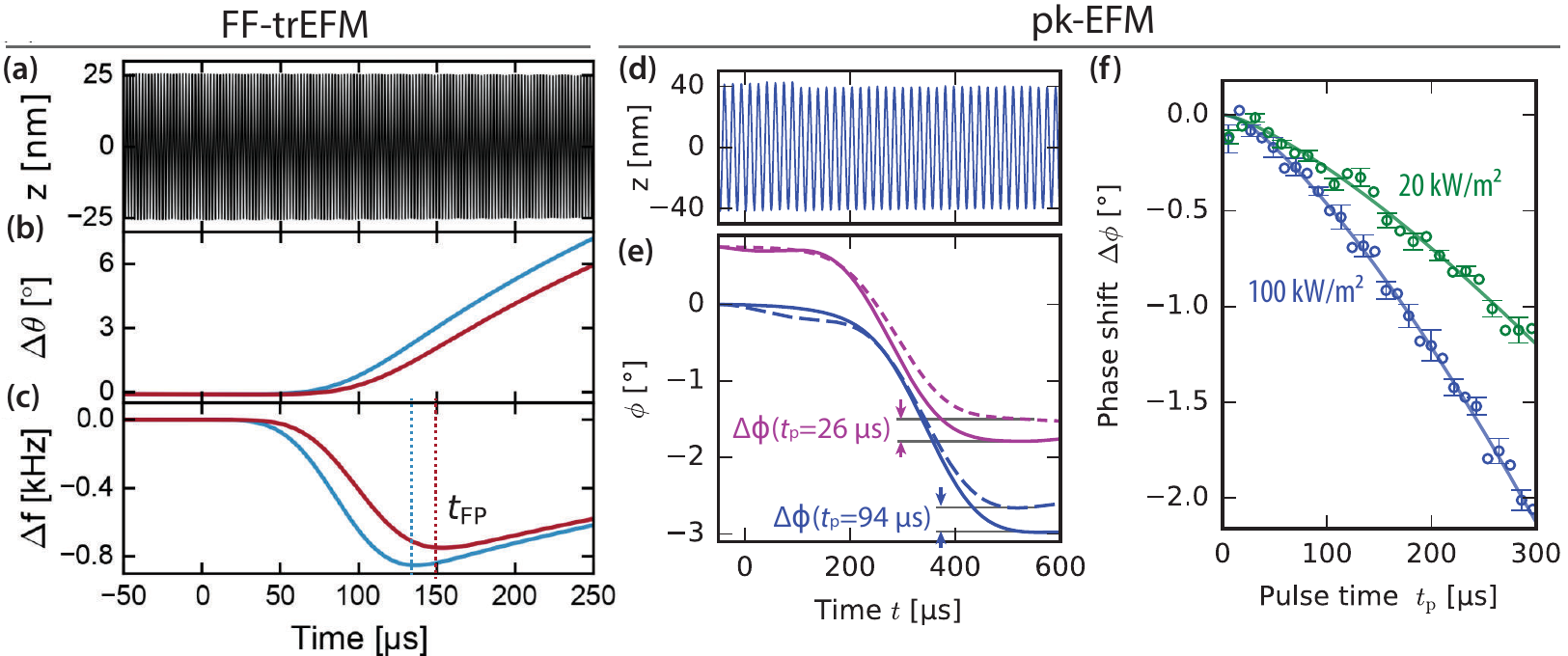}
\caption{FF-trEFM and pk-EFM data. In the FF-trEFM experiment, (a) the cantilever displacement $z$ versus time is digitized and demodulated to produce (b) the cantilever phase shift $\Delta \phi$ and (c) frequency shift $\Delta f$ \latin{versus} time.
Abrupt changes in the tip-sample interaction are inferred from  changes in the time-to-first-peak $t\st{FP}$.
The data are from a control experiment using a voltage pulse with risetime $\tau = \SI{10}{\ns}$ (blue) and $\SI{10}{\us}$ (red).
In the pk-EFM experiment, (d) the cantilever displacement $z$ versus time is digitized and demodulated to produce (e) the cantilever phase shift $\phi$ versus time.
The solid traces are from $t\st{p} = \SI{94}{\us}$ (blue) and $t\st{p} = \SI{26}{\us}$ (purple) experiments; the dashed lines show the corresponding control experiments with the same pulse time but with no light pulse. For each pulse time, the total light-induced phase shift during the pulse $\Delta \phi$ is determined.
Abrupt changes in the tip-sample interaction are inferred from (f) the light-induced phase shift $\Delta \phi$ \latin{versus} pulse time.
Circles show the average phase shift from 6 consecutive pulse times, bars show the standard error, and the lines are a fit to a biexponential model.
Figure (a-c) reproduced with permission from Ref.~\citenum{Karatay2016may} (copyright 2016, American Institute of Physics).
Figure (f) adapted from Ref.~\citenum{Dwyer2017jun} (Creative
Commons Attribution NonCommercial License 4.0, American Association for the Advancement of Science.).
}
\label{fig:trEFM-and-pkEFM-data}
\end{figure*}

How the sample's photo-capacitance dynamics are inferred from the data differs in the two experiments.
In the FF-trEFM experiment the voltage is left on continuously during the measurement; the cantilever oscillation is demodulated to obtain a plot of the cantilever phase and frequency shift versus time (Fig.~\ref{fig:trEFM-and-pkEFM-data}(a-c)).
The transient frequency shift is observed to peak and this time-to-first-peak $t\st{FP}$, Fig.~\ref{fig:trEFM-and-pkEFM-data}(c), can be empirically related back to the photocapacitance rise time $\tau\st{s}$ if suitable control experiments are carried out.
In the pk-EFM experiment the voltage is turned to zero abruptly at a time $t\st{p}$ after the light is turned on.
The cantilever oscillation is again demodulated but instead of studying the transient phase or frequency shift, we measure the light-induced phase shift as a function of the delay time $t\st{p}$ (Fig.~\ref{fig:trEFM-and-pkEFM-data}(d,e)).
Representative data is shown in Fig.~\ref{fig:trEFM-and-pkEFM-data}(f).
In Ref.~\citenum{Dwyer2017jun}, the Fig.~\ref{fig:trEFM-and-pkEFM-data}(f) data was analyzed to reveal that the sample's photocapacitance had biexponential dynamics.

Treating the effect of a time-dependent force $F\st{ts}$ and force gradient $\Delta k$ on cantilever position and momentum is challenging, particularly in the case of photovoltaic materials in which $\tau\st{s}$ can be shorter than the cantilever's period of oscillation.
Nevertheless, using the Lagrangian formalism in conjunction with the Magnus expansion, below we obtain closed-form analytical results for both $t\st{FP}$ and $\Delta \phi(t\st{p})$.
For the phase shift in the pk-EFM experiment, below in Sec.~\ref{Sec:FF-trEFM} we obtain
\begin{equation}
\Delta \phi(t\st{p}) = 
\frac{\Delta C'_{h\nu} V^2 }{2 A_0 k_1}  \frac{\omega_1}{1 + \tau\st{s}^2 \omega_1^2}
\left (
t\st{p} - \tau\st{s} + \tau\st{s} e^{-t\st{p}/\tau\st{s}}
\right ),
\end{equation}
with $A_0$ the cantilever amplitude and $k_1$ and $\omega_1$ the cantilever spring constant and resonance frequency, respectively, in the presence of light and tip voltage.
By fitting the $\Delta \phi$ \latin{versus} $t\st{p}$ data, we can extract both $\Delta C'_{h\nu}$ and $\tau\st{s}$.
The corresponding analytical result for $t\st{FP}$ is more involved; see Eqs.~(\ref{eq:tFP-I}) and (\ref{eq:tFP-II}) below.
The $t\st{FP}$ number obtained in the FF-trEFM experiment depends on $\Delta C'_{h\nu}$, $\Delta C''_{h\nu}$, the cantilever's intrinsic dissipation constant $\gamma$, and $\tau\st{s}$.
Consequently, the time $\tau\st{s}$ cannot be uniquely determined from the single number $t\st{FP}$ measured in the FF-trEFM experiment.
In the pk-EFM experiment, in contrast, the $\Delta \phi$ \latin{versus} $t\st{p}$ data set reveals the full time dependence of the photocapacitance.
Our analysis reveals that the standard equation for frequency shift in KPFM (Eq.~(\ref{eq:df-meas})) cannot be used to analyze these single-shot, transient EFM experiments because the abrupt changes in the tip-sample force shift the cantilever's amplitude and phase.

\begin{figure*}
\includegraphics{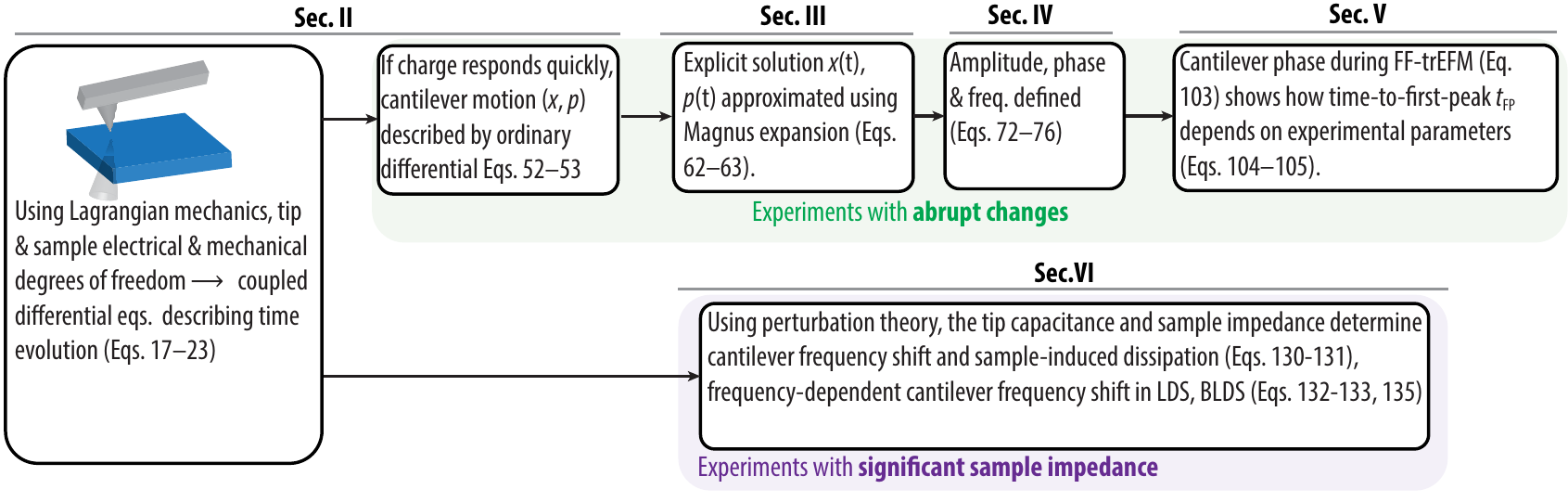}
\caption{Unified Lagrangian-mechanics treatment of electric force microscopy (EFM): an outline showing the structure of the paper, highlighting major results.
}
\label{fig:overview}
\end{figure*}

Figure~\ref{fig:overview} outlines the remainder of the manuscript.
Our overall goal is to explain the results of experiments that violate the assumptions that (1) the tip charge follows the cantilever oscillation instantaneously, and (2) any changes in the tip-sample force or force gradient happen slowly.
In Sec.~\ref{Sec:Lagrangian-introduced} we outline the common Lagrangian formalism that generates coupled differential equations governing the cantilever displacement, tip charge, and any other charges necessary to describe the sample and wiring.
We derive linearized versions of these equations that we then use to treat a variety of EFM experiments.
In Sec.~\ref{Sec:Magnus} we develop an approximate, Magnus-expansion treatment of cantilever dynamics that accurately describes the cantilever position in the event that the tip-sample force and force gradient change abruptly (violating Assumption (2)).
This treatment allows us to describe the phasekick electric force microscopy (pk-EFM) experiment of Ref.~\citenum{Dwyer2017jun} and the feedback-free time-resolved electric force microscopy (FF-trEFM) experiment of Ref.~\citenum{Karatay2016may} using a common formalism (Figs.~\ref{fig:trEFM-and-pkEFM-schematic} and \ref{fig:trEFM-and-pkEFM-data}).
The experimental observables are the cantilever amplitude, phase and frequency, so in Sec.~\ref{sec:Defn-amp-phase} we define these variables in a way that accounts for abrupt changes in the tip-sample force.
The FF-trEFM experiment is discussed in detail in Sec.~\ref{Sec:FF-trEFM}; here we use both the analytical results of Sec.~\ref{Sec:Magnus} and numerical simulations to evaluate the time resolution of the method.
In Sec.~\ref{sec:Impedance_spectroscopy_EFM_theory} we return to the Lagrangian formalism and develop a perturbation-theory approximation that accurately describes the cantilever position and tip charge in the event that the tip charge does not follow the cantilever oscillation instantaneously (violating Assumption (1)).
This approximation describes the cantilever frequency shift and dissipation for an arbitrary sample impedance (Fig.~\ref{fig:Tirmzi2017jan-expt}).
This approximation similarly describes frequency shifts measured in local dielectric spectroscopy \cite{Crider2007jul,Tirmzi2017jan} and broadband local dielectric spectroscopy \cite{Labardi2016may,Tirmzi2017jan} for an arbitrary sample impedance (Fig.~\ref{fig:BLDS-intro}).

The Lagrangian approach to understanding electric force microscopy presented here unifies and significantly expands the treatment of frequency-shift and dissipation effects in EFM presented by Tirmzi \emph{et al.} \cite{Tirmzi2017jan} and Dwyer \emph{et al.} \cite{Dwyer2017jun}.
This approach has a number of advantages.
It accounts for dissipation of energy in both the sample and the cantilever;
treats both steady-state and transient phenomena in a unified way;
incorporates linearization of the equations of motion as an explicit approximation late in the derivation;
and captures the effects of sub-cycle changes in sample capacitance, conductivity, and tip charge that are missing from previous treatments of the cantilever-sample interaction in EFM.
We close by outlining potential avenues of further study in Sec.~\ref{Sec:Discussion}. 

 \section{Cantilever dynamics and tip-sample coupling}
\label{Sec:Lagrangian-introduced}
In this section, we present a general Lagrangian approach for obtaining coupled equations of motion for the EFM cantilever, tip-sample charge, and external tip-sample bias circuitry. The EFM cantilever, sample, and bias circuitry constitute a coupled electro-mechanical system of the type considered by Wells \cite{Wells1938may,Wells1967}, Ogar \cite{Ogar1962mar}, and others \cite{Meisel1966,Sira-Ramirez1996may,Weiss1997sep,Aspelmeyer2014}. These authors demonstrate that the equations of motion for such systems can be developed in a unified Lagrangian formalism with the electrical behavior treated in the lumped circuit element approximation of elementary circuit theory. In our analysis, the electrical behavior of the sample is modeled by a single complex impedance, while the tip-sample coupling is modeled as a position-dependent capacitance $C\stt{T}$, with charge $q\stt{T}$, connected in series with the sample impedance. The complete circuit, consisting of tip, sample, and external bias, could be analyzed by applications of Kirchhoff's junction rule and loop rule; however we find it advantageous to take the Lagrangian approach, described in detail below, as the correct electro-mechanical coupling terms arise naturally in a unified framework.

\begin{figure}
\includegraphics{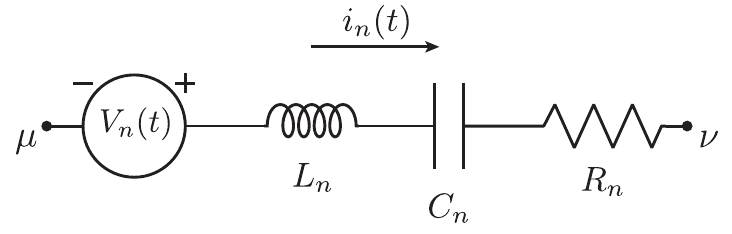}
\caption{The $n^{\mathrm{th}}$ branch of the circuit between junctions $\mu$ and $\nu$ with instantaneous current $i_n(t)$. The circuit branch behaves as a voltage source $V_n(t)$, inductance $L_n$, capacitance $C_n$, and resistance $R_n$ connected in series, while its state is specified by the generalized coordinate $q_n$ and its time derivative $\dot{q}_n=i_n(t)$. Taking $V_n=0$, $L_n=0$, $1/C_n=0$, or $R_n=0$ is equivalent to omitting the corresponding circuit element. Each junction $\alpha$ in the circuit is characterized by a set of branch currents $\{\mathrm{in}_{\alpha}\}$ directed into the junction and another set of branch currents $\{\mathrm{out}_{\alpha}\}$ directed out of the junction. Based on the direction of the current indicated in the illustration, $i_n\in\{\mathrm{out}_{\mu}\}$ and $i_n\in\{\mathrm{in}_{\nu}\}$.}\label{Fig:branch}
\end{figure}

The circuit representing the electrical degrees of freedom of the EFM consists of branches---discrete circuit elements wired in series as illustrated in \figref{Fig:branch}---interconnected by electrical junctions at each end.
For notational purposes, each circuit branch is identified by a Latin subscript (e.g., $n$ in \figref{Fig:branch}), while each junction is identified by a Greek subscript (e.g., $\mu$ and $\nu$ in \figref{Fig:branch}).
Specification of the circuit branches, their interconnections, the cantilever mechanical properties, and the position-dependent tip-sample capacitance constitutes the complete model. 

The Lagrangian and Rayleigh dissipation function of the EFM have contributions arising from the circuit branches, the circuit junctions, and the mechanical degrees of freedom. In the following treatment, we identify contributions from the circuit branches with the subscript $\mbox{B}$, contributions from the circuit junctions with the subscript $\mbox{J}$, and contributions from the mechanical degrees of freedom by the subscript $\mbox{M}$.  

The generalized coordinate specifying the state of the $n^{\mathrm{th}}$ circuit branch is
\begin{equation}
q_n=q_{n0}+\int_{t_0}^t i_n(t')dt',
\label{Eq:branch-charge}
\end{equation}
where $q_{n0}$ is the charge $q_{n}$ at the initial time $t_0$. Then
\begin{equation}
\dot{q}_n=i_n(t)
\label{Eq:branch-current}
\end{equation}
is the instantaneous current through the $n^{\mathrm{th}}$ branch. Collectively, the branches of the circuit, as shown in Fig.~\ref{Fig:branch}, contribute the additive terms
\begin{equation}
\mathcal{L}\stt{B}=\sum_n\frac{L_n \dot{q}_n^2}{2}-\frac{q_n^2}{2C_n}+V_nq_n
\label{Eq:branch-Lagrangian}
\end{equation}
to the Lagrangian and
\begin{equation}
\mathcal{D}\stt{B}=\sum_n\frac{R_n\dot{q_n}^2}{2}
\label{Eq:branch-dissipation}
\end{equation}
to the Rayleigh dissipation function when the corresponding circuit elements are present.

\begin{figure*}
\includegraphics{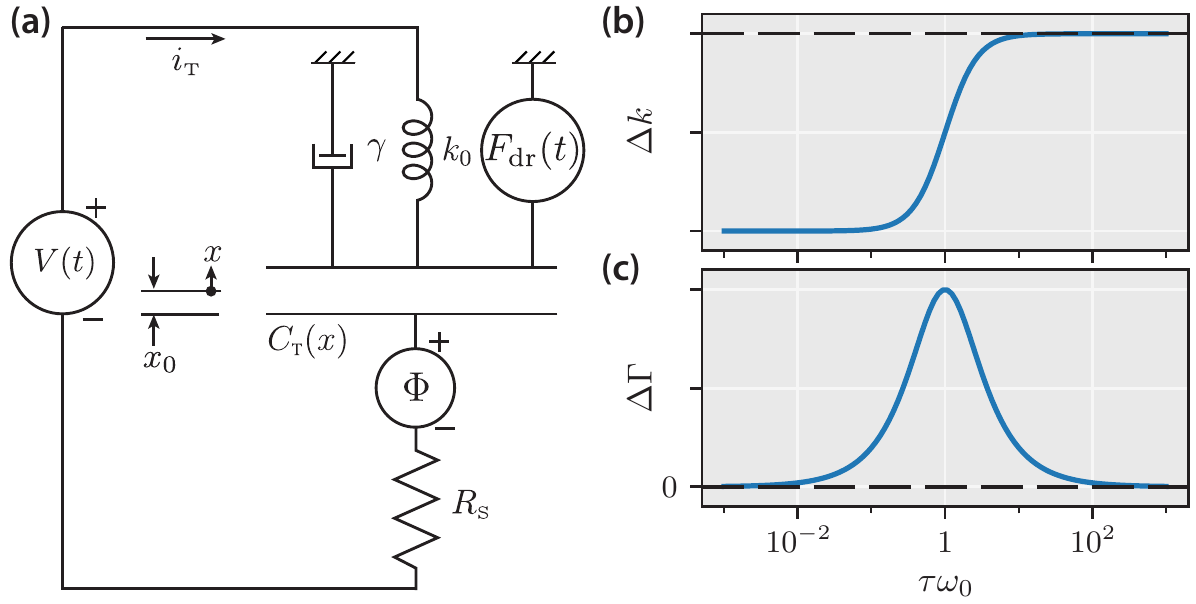}
\caption{(a) Single-loop circuit model of an EFM experiment with a resistive sample.
Changes in the effective cantilever frequency and damping constant depend on the cantilever resonance frequency $\omega_0$ and the time constant $\tau$ as defined in Eq.~(\ref{Eq:dissipation-linear-q}).
(b) Changes to the effective cantilever spring constant due to the tip-sample interaction.  
For $\tau \ll 1 / \omega_0$, oscillating charge reduces the effective spring constant, while for $\tau \gg 1 / \omega_0$, this effect is suppressed. (c) Additional cantilever damping due to the tip-sample interaction. The damping is maximized for $\tau = 1 / \omega_0$.}
\label{Fig:current-induced-dissipation-circuit}
\end{figure*}

The constraints of charge conservation at the circuit junctions are fully realized by the application of Kirchhoff's junction rule to a set ${\left\{\alpha\right\}}$ of all but one of the junctions, each of which is characterized by sets $\{\mathrm{out}_{\alpha}\}$ and $\{\mathrm{in}_{\alpha}\}$ of outward and inward directed branch currents (see, \latin{e.g.}, Ref.~\onlinecite{Mayergoyz2012dec}). In our analysis, we enforce these constraints via the method of Lagrange multipliers, adding the term
\begin{equation}
\mathcal{L}\stt{J}=\sum_{\alpha}\lambda_{\alpha} \left(\sum_{i_n\in\{\mathrm{out}_{\alpha}\}} q_n - \sum_{i_n\in\{\mathrm{in}_{\alpha}\}} q_n\right)
\label{Eq:junction-Lagrangian}
\end{equation}
to the system's Lagrangian.  The Lagrange multipliers $\lambda_{\alpha}$ are then treated as additional generalized coordinates of the system, which, when defined as in \eqnref{Eq:junction-Lagrangian}, can be identified with the instantaneous electric potential of the associated junctions referenced to the one omitted junction.

The mechanical degrees of freedom of the cantilever and all tip-sample forces $F_0$ not arising from capacitive coupling are included with the usual Lagrangian $\mathcal{L}\stt{M}$, Rayleigh dissipation function $\mathcal{D}\stt{M}$ and generalized forces $\mathcal{F}_x$. In keeping with the conventional notation for one-dimensional harmonic oscillators and to avoid confusion with the convention of using $z$ to represent a complex number, for the remainder of the article we represent cantilever displacement with $x$ rather than $z$, with increasing $x$ corresponding to motion of the cantilever tip away from the sample surface.

Having accounted for all relevant degrees of freedom, we generate the coupled electro-mechanical equations of motion by application of the Euler-Lagrange equation
\begin{equation}
\frac{d}{dt} \left( \frac{\partial {\cal L}}{\partial \dot{q}_n} \right)
 - \frac{\partial {\cal L}}{\partial q_n} 
 = - \frac{\partial {\cal D}}{\partial \dot{q}_n} + {\cal F}_n
 \label{Eq:Euler-Lagrange}
\end{equation}
to each generalized coordinate $q_n$, where
\begin{equation}
\mathcal{L}=\mathcal{L}\stt{B}+\mathcal{L}\stt{J}+\mathcal{L}\stt{M}
\label{Eq:Lagrangian}
\end{equation}
and
\begin{equation}
\mathcal{D}=\mathcal{D}\stt{B}+\mathcal{D}\stt{M}.
\label{Eq:dissipation}
\end{equation}
Note that in writing \eqnref{Eq:Euler-Lagrange}, we have extended the range of the index $n$ and understand $x$ and the $\lambda_\alpha$'s to be among the generalized coordinates $q_n$.

In all of the cases we consider, the mechanical EFM cantilever is modeled as a linear harmonic oscillator with mass $m$, spring constant $k_0=m\omega_0^2$, linear damping coefficient $\Gamma=2m\gamma$, and applied drive force $F\st{dr}(t)$, giving
\begin{equation}
\mathcal{L}\stt{M}=\frac{m\dot{x}^2}{2}-\frac{m\omega_0^2x^2}{2},
\label{Eq:L-M}
\end{equation}
\begin{equation}
\mathcal{D}\stt{M} = m\gamma\dot{x}^2,
\label{Eq:D-M}
\end{equation}
and
\begin{equation}
\mathcal{F}_x=F_0(x)+F\st{dr}(t).
\label{Eq:F-x}
\end{equation}
Using \eqnsref{Eq:L-M}--(\ref{Eq:F-x}), and noting that
\begin{equation}
\pd{\mathcal{L}}{x}
=\frac{ C'\stt{T}(x) q\stt{T}^{2}}{2C\stt{T}(x)^2}
\label{Eq:dLBdx}
\end{equation}
irrespective of the bias circuitry or sample impedance, application of \eqnref{Eq:Euler-Lagrange} for the generalized coordinate $x$ gives
\begin{equation}
m\ddot{x}+2m\gamma\dot{x}+m\omega^{2}_{0}x-\frac{ C'\stt{T}(x) q\stt{T}^{2}}{2C\stt{T}(x)^2}=F_0(x)+F\st{dr}(t).
\label{Eq:general-x-eom}
\end{equation}
Throughout the article we determine the cantilever displacement by solving or approximating \eqnref{Eq:general-x-eom}. The equations of motion of the charge degrees of freedom to which \eqnref{Eq:general-x-eom} is coupled, on the other hand, vary from model to model.

The capacitive coupling and $F_0(x)$ terms that comprise the tip-sample force in \eqnref{Eq:general-x-eom} are nonlinear in general. The nonlinearity of $F_0(x)$ is of particular concern in high-resolution AFM imaging where it has been shown to cause significant amplitude dependence of the cantilever oscillation frequency \cite{Giessibl1997dec, Holscher1999, Garcia2002sep} and lead to bi-stability in driven cantilevers with amplitude feedback control of the tip-sample separation \cite{Garcia1999aug, Garcia2000may, Garcia2002sep}. The EFM experiments that we consider involve minimum tip-sample separations of 10's of nanometers that are beyond the effective range of the nonlinearities in $F_0(x)$ \cite{Holscher1999}. The approach in the following analysis is to neglect $F_0(x)$ and to solve small-amplitude linearized approximations of the resulting EFM equations of motion. These approximations are not too severe in that this approach is sufficient to explain the data in the experiments of Figs.~\ref{fig:Tirmzi2017jan-expt}--\ref{fig:trEFM-and-pkEFM-data}. We defer further discussion of the significance of the small amplitude approximation and neglecting the nonlinearities of \eqnref{Eq:general-x-eom} until Sec.~\ref{Sec:Discussion}.
With the general theory completely developed, we proceed to the characterization of specific EFM experiments.

\subsection{Current-induced cantilever dissipation}
\label{Sec:Current-induced-dissipation}
In this section we apply the Lagrangian theory to a simple model that violates the assumption of the tip charge following the cantilever oscillation instantaneously.
In this model, as shown in Fig.~\ref{Fig:current-induced-dissipation-circuit}, a voltage $V(t)$ is applied between the cantilever tip and sample, while the tip displacement $x$ changes the tip-sample capacitance $C\stt{T}(x)$.
The surface potential is represented by the voltage source $\Phi$ and the sample has a resistance $R\stt{S}$.
By inspection, the branch Lagrangian and dissipation function are
\begin{equation}
\mathcal{L}\stt{B}=-\frac{q\stt{T}^2}{2C\stt{T}(x)}+(V(t) - \Phi) q\stt{T},
\label{Eq:dissipation-branch-L}
\end{equation}
and
\begin{equation}
\mathcal{D}\stt{B}=\frac{R\stt{S}\dot{q}\stt{T}^2}{2}.
\label{Eq:dissipation-branch-D}
\end{equation}
As the circuit consists of a single branch, $\mathcal{L}\stt{J}=0$.
We generate the equations of motion by applying the Euler-Lagrange equation (\eqnref{Eq:Euler-Lagrange}).
The equation of motion for the tip displacement is given by \eqnref{Eq:general-x-eom} with $F_0(x)=0$.
The equation of motion for the tip charge is
\begin{equation}
R\stt{S}\dot{q}\stt{T}= (V(t)-\Phi) - \frac{q\stt{T}}{C\stt{T}(x)}.
\label{eq:qT-eom-Rs}
\end{equation}

We now show that the simple model of Fig.~\ref{Fig:current-induced-dissipation-circuit} is sufficient to reproduce the characteristic cantilever dissipation seen in EFM experiments such as those described in Refs.~\onlinecite{Denk1991oct, Cockins2010may, Zhu2008aug}.
In particular, cantilever dissipation is proportional to $V^2$ and $C'^2$ and is maximized when the tip charging rate matches the cantilever frequency.
In the experiment of Denk and Pohl \cite{Denk1991oct}, an external drive force $F\st{dr}$ induces a small oscillation at the cantilever's resonance frequency.
Small changes in the cantilever's resonance frequency and dissipation are measured as a function of the static applied voltage.
To model this experiment, we seek solutions to the above system of coupled nonlinear differential equations in the form of small driven oscillations about the equilibrium point $x_0$ and tip charge $q_0$.
To this end, we expand the tip-sample capacitance to second order about $x_0$, giving
\begin{equation}
C\stt{T}(x) \approx C_0 + C'(x-x_0)+\frac{1}{2}C''(x-x_0)^2,
\label{Eq:C-expand}
\end{equation}
and then make the change of variables
\begin{equation}
\begin{aligned}
x &\rightarrow x+x_0\\
q\stt{T} &\rightarrow q\stt{T}+q_0
\end{aligned}
\label{Eq:change-of-variable}
\end{equation}
so that $x$ and $q\stt{T}$ now represent a small change from the equilibrium point.
The linearized equations of motion are
\begin{equation}
m\ddot{x}+2m\gamma\dot{x}+m\omega^{2}_{0}x- \frac{C'' V^2}{2}x+\frac{C'V}{C_0}\tau \dot{q}\stt{T}=F\st{dr}(t),
\label{Eq:dissipation-linear-x}
\end{equation}
and 
\begin{equation}
\tau\dot{q}\stt{T}= C'Vx - q\stt{T},
\label{Eq:dissipation-linear-q}
\end{equation}
where $\tau=R\stt{S}C_0$ is the tip-sample charging time constant and we combine the applied voltage and surface potential as $V=V(t)-\Phi$ for notational efficiency.
\footnote{For the reader interested in deriving \eqnref{Eq:dissipation-linear-x} and \eqnref{Eq:dissipation-linear-q}, it is helpful to begin by using \eqnref{eq:qT-eom-Rs} to rewrite $q\stt{T}^2/C\stt{T}(x)^2$ in \eqnref{Eq:general-x-eom} as $\left(V-R\stt{S}\dot{q}\stt{T}\right)^2$ and then to multiply \eqnref{eq:qT-eom-Rs} through by $C\stt{T}(x)$ before proceeding to expand $C\stt{T}(x)$ about $x_0$. The linearized equations follow from applying the equilibrium condition to identify $x_0$ and $q_0$, making the substitutions indicated in \eqnref{Eq:change-of-variable}, and dropping terms that are nonlinear in the new coordinates and their time derivatives. Note that the expressions for $x_0$ and $q_0$ do not rely on approximating $C\stt{T}(x)$ in a power series and that the coefficients on the right hand side of \eqnref{Eq:C-expand} are to be evaluated at the equilibrium position of the cantilever under the tip-sample interaction, not at the equilibrium position of the non-interacting cantilever.
}

We now consider the steady-state solution when the cantilever is subject to 
\begin{equation}
F\st{dr}(t)= \Re{F(\omega) e^{j\omega t}},
\end{equation}
where $F(\omega)$ is the complex amplitude of the oscillating driving force.
In the linear-response regime the position and tip charge have the form
\begin{equation}
x(t)=\Re{x(\omega)e^{j\omega t}},
\label{Eq:dissipation-oscillating-position}
\end{equation}
and
\begin{equation}
q\stt{T}(t)=\Re{q\stt{T}(\omega)e^{j\omega t}}.
\label{Eq:dissipation-oscillating-charge}
\end{equation}
Substituting Eqs.~(\ref{Eq:dissipation-oscillating-position}) and (\ref{Eq:dissipation-oscillating-charge}) into Eqs.~(\ref{Eq:dissipation-linear-x}) and (\ref{Eq:dissipation-linear-q}) gives
\begin{multline}
\left(
-m\omega^2
+j\omega2m\gamma
+m\omega^{2}_{0}
\vphantom{\frac{V^2C''}{2}} \right. \\
\left.
\underbrace{-\frac{C'' V^2}{2}+\frac{C'^2 {V}^2}{C_0}\frac{j\omega \tau} {1+j\omega\tau}}\right) x(\omega)
=F(\omega).
\label{Eq:dissipation-frequency-response}
\end{multline}
This equation has the form
\begin{equation}
\Big(
{- m\omega^2}
+j\omega(\Gamma+\Delta\Gamma)
+(k_{0}+\Delta k)\Big)
x(\omega)
=F(\omega),
\label{Eq:damped-HO-frequency-response}
\end{equation}
which describes the response of a damped harmonic oscillator with additional damping
\begin{equation}
\Delta\Gamma=2m\Delta\gamma=\frac{V^2C'^2}{\omega C_0}\frac{\omega\tau}{1+\omega^2\tau^2}
\label{Eq:dissipation-additional-damping}
\end{equation}
and additional spring constant  
\begin{equation}
\Delta k = -\frac{V^2C''}{2}+\frac{V^2C'^2}{C_0}\frac{\omega^2\tau^2}{1+\omega^2\tau^2}
\label{Eq:dissipation-additional-k}
\end{equation}
arising from the imaginary and the real parts of the under-braced term in Eq.~(\ref{Eq:dissipation-frequency-response}) respectively.
In the limit that $\tau \rightarrow 0$, Eq.~\ref{Eq:dissipation-additional-k} recovers the simplified Eq.~\ref{eq:Df-SKPM-const-V} behavior.
In Ref.~\citenum{Tirmzi2017jan}, Eq.~\ref{Eq:dissipation-additional-k} was used to analyze the observed frequency shift.

In the approximation that the tip-sample interaction can be modeled as a parallel-plate capacitor, Eq.~(\ref{Eq:dissipation-additional-k}) takes on a particularly simple form.
In this approximation $C'^2/C_0=C''/2$, and the additional spring constant shift simplifies to
\begin{equation}
\Delta k = -\frac{V^2 C''}{2} \frac{1}{1+\omega^2\tau^2}.
\label{Eq:dissipation-additional-k-II}
\end{equation}
In the parallel-plate case, when $\tau \rightarrow \infty$, $\Delta k \rightarrow 0$.
In a scanned probe experiment, the parallel-plate model is a poor description of the tip-sample interaction; in this case, $\Delta k$ in the $\tau \rightarrow \infty$ limit is nonzero, and depends on $C^{\prime\prime}$, $C^{\prime}$, and $C_0$.

Equation (\ref{Eq:dissipation-additional-damping}) demonstrates the expected $\Delta\Gamma\propto V^2C'^2$ behavior.
Figure~\ref{Fig:current-induced-dissipation-circuit}(b) and (c) illustrate the behavior of $\Delta \Gamma$ and $\Delta k$ as $\tau$ varies from the fast-charging limit to the slow-charging limit while the cantilever is driven at its resonance frequency.
For fast charging ($\tau \omega_0 \ll 1$), $q\stt{T}$ oscillates in phase with the cantilever and there is no additional damping.
As $\tau$ increases, $q\stt{T}$ begins to oscillate out of phase with the cantilever leading to an increase in $\Delta \Gamma$ that peaks as expected at $\tau=\omega_0^{-1}$.
For slow charging, with $\tau$ much longer than the cantilever period, $q\stt{T}$ no longer oscillates significantly and the additional dissipation vanishes. This dependence of cantilever damping on the charging rate agrees with previous results. For example, Miyahara \latin{et al}.\ present the same dependence derived in the context of cantilever-induced single-electron tunneling \cite{Miyahara2017jan}.  

While this single-loop circuit model captures the essential physics of cantilever damping and frequency shifts due to finite $\tau$, it neglects many potentially significant features of real experiments, such as stray capacitance and resistance in the external wiring and complex impedance of the sample. We proceed by treating these features in the Lagrangian formalism to develop equations of motion that apply to a wide range of EFM protocols, returning to address the case of non-negligible sample impedance in detail in Sec.~\ref{sec:Impedance_spectroscopy_EFM_theory}.

\subsection{A more general EFM model}
\label{Sec:more-realistic-EFM}

\begin{figure}
\includegraphics{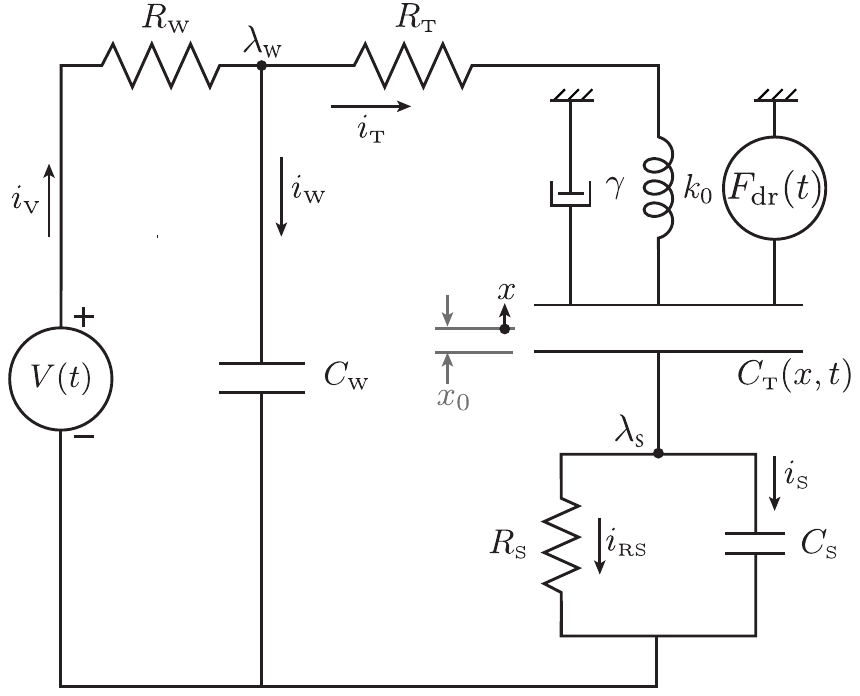}
\caption{Equivalent circuit of a generalized EFM experiment.
This circuit accounts for stray capacitance $C\stt{W}$ and resistance $R\stt{W}$ in the external wiring, resistance $R\stt{T}$ in the cantilever tip, tip-sample capacitance $C\stt{T}$, sample capacitance $C\stt{S}$, and sample resistance $R\stt{S}$.}  
\label{Fig:photocapacitance-EFM-circuit}
\end{figure}

Figure~\ref{Fig:photocapacitance-EFM-circuit} illustrates our generalized EFM model, which accounts for the applied bias $V(t)$, resistance $R\stt{W}$ and capacitance $C\stt{W}$ in the external wiring, resistance $R\stt{S}$ and capacitance $C\stt{S}$ of the sample, resistance $R\stt{T}$ between the cantilever base and tip, and the tip-sample capacitance $C\stt{T}(x,t)$. Again, the surface potential is incorporated into $V(t)$ for notational convenience. In addition to the position dependence of $C\stt{T}$, which couples the electrical and mechanical degrees of freedom of the EFM, we consider the possibility that $C\stt{T}$ is explicitly time dependent as is the case in, \latin{e.g.}, photocapacitance measurements. Applying Eqs.~(\ref{Eq:branch-Lagrangian})--(\ref{Eq:junction-Lagrangian}) to the circuit of \figref{Fig:photocapacitance-EFM-circuit}, we have
\begin{equation}
\mathcal{L}\stt{B}=-\frac{q\stt{W}^2}{2C\stt{W}}-\frac{q\stt{T}^2}{2C\stt{T}(x,t)}-\frac{q\stt{S}^2}{2C\stt{S}}+V(t)q\stt{V},
\label{Eq:L-B}
\end{equation}
\begin{equation}
\mathcal{D}\stt{B}=\frac{R\stt{W}\dot{q}\stt{V}^2}{2}+\frac{R\stt{T}\dot{q}\stt{T}^2}{2}+\frac{R\stt{S}\dot{q}\stt{RS}^2}{2},
\label{Eq:D-B}
\end{equation}
and
\begin{equation}
\mathcal{L}\stt{J}=\lambda\stt{W}(q\stt{W}+q\stt{T}-q\stt{V})+\lambda\stt{S}(q\stt{RS}+q\stt{S}-q\stt{T}).
\label{Eq:L-J}
\end{equation}

In total, there are two Lagrange multipliers, five branch coordinates, and one mechanical coordinate, requiring eight applications of \eqnref{Eq:Euler-Lagrange} to generate the equations of motion. The time derivatives of the two equations generated by the Lagrange multipliers simply reproduce the junction-rule relations
\begin{subequations}
\begin{equation}
\dot{q}\stt{V}=\dot{q}\stt{T}+\dot{q}\stt{W}
\end{equation}    
and
\begin{equation}
\dot{q}\stt{RS}=\dot{q}\stt{T}-\dot{q}\stt{S}.
\end{equation}
\label{Eq:junction-eoms}
\end{subequations}
The equations generated by the branch coordinates $q\stt{W}$ and $q\stt{S}$,
\begin{subequations}
\begin{equation}
\lambda\stt{W}=q\stt{W}/C\stt{W}
\end{equation}    
and
\label{Eq:multiplier-eoms}
\begin{equation}
\lambda\stt{S}=q\stt{S}/C\stt{S},
\end{equation}
\end{subequations}
give algebraic expressions for the Lagrange multipliers and confirm the earlier assertion about their relationship to the electric potential.

Using \eqnsref{Eq:junction-eoms} and (\ref{Eq:multiplier-eoms}) to eliminate $q\stt{V}$, $q\stt{RS}$, $\lambda\stt{W}$ and $\lambda\stt{S}$, the remaining four equations of motion can be written as
\begin{equation}
m\ddot{x}+2m\gamma\dot{x}+m\omega^{2}_{0}x-\frac{C\stt{T}'(x)q\stt{T}^{2}}{2C^2\stt{T}}=F\st{dr}(t),
\label{Eq:x-eom}
\end{equation}
\begin{equation}
R\stt{T}\dot{q}\stt{T}=\frac{q\stt{W}}{C\stt{W}}-\frac{q\stt{T}}{C\stt{T}}-\frac{q\stt{S}}{C\stt{S}},
\label{Eq:q_T-eom}
\end{equation}
\begin{equation}
R\stt{W}\left(\dot{q}\stt{W}+\dot{q}\stt{T}\right)=V(t)-\frac{q\stt{W}}{C\stt{W}},
\label{Eq:q_W-eom}
\end{equation}
and
\begin{equation}
R\stt{S}\left(\dot{q}\stt{T}-\dot{q}\stt{S}\right)=\frac{q\stt{S}}{C\stt{S}}.
\label{Eq:q_S-eom}
\end{equation}
These four equations represent a complete model for a broad class of EFM experiments.
As we show in the next section, significant simplifications to this system of equations can be realized in experiments characterized by fast charging and small oscillations.

\subsection{Cantilever dynamic in the fast charging, small oscillation limit}
\label{Sec:Lagrangian-fast-small-limit}

For many EFM experiments, including the photocapacitance measurements described in Secs.~\ref{Sec:Magnus}--\ref{Sec:FF-trEFM}, the capacitive charge redistribution times are much faster than one cantilever cycle and voltage drops across the resistances (\latin{i.e.}, the left hand sides of Eqs.~(\ref{Eq:q_T-eom}--\ref{Eq:q_S-eom})) are negligible. Taking the resistances in the equations of motion to zero independently implements this fast-charging limit.  In particular, taking $R_{\mbox{\tiny{S}}}\rightarrow0$ in \eqnref{Eq:q_S-eom} implies
\begin{equation}
q\stt{S}=0,
\label{Eq:fast-S}
\end{equation}
while taking $R\stt{W}\rightarrow0$ in \eqnref{Eq:q_W-eom} implies
\begin{equation}
\frac{q\stt{W}}{C\stt{W}}=V(t),
\label{Eq:fast-W}
\end{equation}
and taking $R\stt{T}\rightarrow0$ in \eqnref{Eq:q_T-eom} implies
\begin{equation}
\frac{q\stt{T}}{C\stt{T}}=\frac{q\stt{W}}{C\stt{W}}-\frac{q\stt{S}}{C\stt{S}}.
\label{Eq:fast-T}
\end{equation}
When all three resistances are negligible, \eqnsref{Eq:fast-S}--(\ref{Eq:fast-T}) require $q\stt{T}/C\stt{T}=V(t)$, or
\begin{equation}
m\ddot{x}+2m\gamma\dot{x}+m\omega^{2}_{0}x-\frac{1}{2}V(t)^{2}\pd{C\stt{T}}{x}=F\st{dr}(t).
\label{Eq:fast-x}
\end{equation}

For sufficiently small cantilever oscillation amplitude, the tip-sample capacitance gradient can be linearized in $x$.  In this approximation, with
\begin{subequations}
\begin{equation}
p=m\dot{x},
\end{equation}
\eqnref{Eq:fast-x} becomes
\label{Eq:linear-fast-x}
\begin{equation}
\dot{p}+2\gamma p+m\omega^{2}_{0}x-\frac{1}{2}V(t)^{2}\left(C'(t)+C''(t)x\right)=F\st{dr}(t).
\end{equation}
\end{subequations}
In \eqnref{Eq:linear-fast-x} we have reduced the equations of motion to a pair of first order ordinary differential equations that govern both the pk-EFM and the FF-trEFM experiments described in Sec. \ref{Sec:Introduction}.
Note that in the FF-trEFM literature, the term $V(t)^2 C''(t)/2$ is accounted for as a time-dependent natural resonance frequency $\omega_0(t)$, which is an important notational difference from our usage where $\omega_0$ is the cantilever resonance frequency in the absence of capacitive coupling between the tip and sample \cite{Giridharagopal2012jan,Karatay2016may}.
In the next section we demonstrate an approximate solution to \eqnref{Eq:linear-fast-x} that is particularly well-suited to describe the cantilever motion in terms of time-dependent frequency and phase shifts.

\section{Magnus expansion treatment of photocapacitance measurements}
\label{Sec:Magnus}

In this section, we develop a Magnus-expansion solution for the cantilever motion during a photocapacitance measurement, extending our previous results from Ref.~\onlinecite{Dwyer2017jun} to include both phasekick electric force microscopy (pk-EFM) and feedback-free time-resolved electric force microscopy (FF-trEFM) experiments in a common formalism.
Equations~(\ref{Eq:linear-fast-x}) are two coupled, linear ordinary differential equations with time-varying coefficients.
Noting that the $C''$ term in \eqnref{Eq:linear-fast-x} gives rise to a shift $\Delta k(t)$ in the effective spring constant, we define the fractional change in spring constant
\begin{equation}
\kappa(t) =
\frac{\Delta k(t)}{k_0} \equiv
-\frac{1}{2m\omega_0^2}C''(t)V(t)^2.
\label{eq:dk-kappa}
\end{equation}
The $C'$ term in \eqnref{Eq:linear-fast-x} is the tip-sample force $F\st{ts}(t) = \frac{1}{2} C' V(t)^2$.
The total force is
\begin{equation}
F(t) =
    F\st{ts}(t)
		+F\st{dr}(t)
    \equiv
		\frac{1}{2} 
    C'(t)
    V(t)^2
		+F\st{dr}(t).
\label{eq:force-sum}
\end{equation}
Using these definitions, \eqnref{Eq:linear-fast-x} can be written in terms of the position-momentum state vector $\bm{x} = (x \, \, p)^T$ as
\begin{align}
\dot{\bm{x}}& = \bm{A}(t) \bm{x} + \bm{b}(t),
\label{eq:general-state-evolution}
\end{align}
where 
\begin{align}
\bm{A}(t)& =
\begin{pmatrix}
0 & 1/m \\
-m \omega_0^2
 \: (1 + \kappa(t)) & -2 \gamma
\end{pmatrix},
\label{eq:A-matrix}
\end{align}
and
\begin{align}
\bm{b}(t) &= 
\begin{pmatrix}
0 \\
F(t)
\end{pmatrix}.
\label{eq:b-vector}
\end{align}

While there is no general analytic solution to \eqnref{eq:general-state-evolution}, we can use the Magnus-expansion technique to obtain a highly accurate approximation \cite{Magnus1954nov, Blanes2009jan}.
The exact solution can be written in terms of the system's (unknown) propagator $\bm{U}$,
\begin{equation}
\bm{x}(t)
    = \bm{U}(t,t_0) \, \bm{x}(t_0)
    + \int_{t_0}^{t}
        \bm{U}(t,t') \,
        \bm{b} (t') \: 
      dt'.
\label{eq:propagator-evolve-state-plus-forcing-term}
\end{equation}
To take the Magnus expansion, we write $\bm{U}(t,t_0)$ as the exponential of a matrix $\bm{\Omega}$:
\begin{equation}
\bm{U}(t, t_0) \equiv \exp{\bm{\Omega}(t, t_0)},
\label{eq:define-Omega}
\end{equation}
and approximate $\bm{U}$ by approximating $\bm{\Omega}$.
The first order Magnus approximation for $\bm{\Omega}$ is 
\begin{align}
\bm{\Omega}(t,t_0) &
    \approx
    \int_{t_0}^{t}
        \bm{A}(t')
      d t'.
\label{eq:x-of-t}
\end{align}
For high-quality-factor cantilevers ($Q \equiv \omega_0/(2\gamma) \gg 1$), matrix exponential can be approximated\footnote{
The matrix exponential is most easily calculated using the eigendecomposition $\bm{\Omega} = \bm{Q}\bm{\Lambda} \bm{Q}^{-1}$, where $\bm{Q}$ is the matrix whose columns are the eigenvectors of $\bm{\Omega}$ and $\bm{\Lambda}$ is the diagonal matrix with the corresponding eigenvalues of $\bm{\Omega}$ along the diagonal. The result in \eqnref{eq:propagator} is obtained by approximating the eigenvalues and eigenvectors to first-order in $\gamma$, a good approximation when $\gamma \ll \bar{\omega}$. The approximate eigenvalues are $-\gamma \pm j\bar{\omega}$ and the approximate eigenvectors are $\big ( -\frac{\gamma}{m \bar{\omega}^2} \pm \frac{1}{j m\bar{\omega}}, 1 \big)^T$.
}
to give a propagator
\begin{widetext}
\begin{equation}
\bm{U}(t,t_0)
\approx
e^{-\gamma \Delta t}
\begin{pmatrix*}[c]
\cos{\left (\, \bar{\omega}
     \, \Delta t \right )}  +\dfrac{ \sin{ \left ( \, \bar{\omega} 
    \, \Delta t \right)} }{2Q}\hphantom{m} & 
\dfrac{\sin{\left( \, \bar{\omega} \, \Delta t \right)}}{m \, \bar{\omega}}\\
- m \, \bar{\omega}
\sin{\left( \, \bar{\omega}
     \, \Delta t \right)} \hphantom{m} &
\cos ( \, \bar{\omega} 
    \, \Delta t) - \dfrac{\sin{\left( \, \bar{\omega}\, \Delta t \right) }}{2Q} 
\end{pmatrix*},
\label{eq:propagator}
\end{equation}
\end{widetext}
where $\Delta t = t-t_0$ is the elapsed time and $\bar{\omega}$ is the \emph{time-dependent} average frequency defined by
\begin{equation}
\bar{\omega} \equiv
    \omega_0 \left(
        1 + \frac{1}{\Delta t} 
           \int_{t_0}^{t} 
              \kappa(t') \: 
              d t'
    \right)^{1/2}.
    \label{eq:propagator-average-frequency}
\end{equation}
If $\kappa(t)$ is constant, then the first order Magnus expansion is exact.
For a time varying $\kappa(t)$, corrections to the exponent $\bm{\Omega}$ will be on the order of the \emph{change} in the normalized spring constant shift $\Delta \kappa$.
An unrealistically large light-induced frequency shift of $\Delta f = \SI{350}{\Hz}$ for a $f_0 = \SI{70}{\kHz}$ resonance frequency cantilever corresponds to a change in the normalized spring constant $\Delta \kappa\st{max} = 2 \Delta f / f_0 = \num{0.01}$.
We are justified in neglecting higher-order terms of the Magnus expansion because $\Delta \kappa\st{max} \ll 1$.

To derive the usual EFM expression for cantilever frequency, we define the cantilever phase accumulated between $t_0$ and $t$: $\theta(t,t_0) \equiv \bar{\omega} \, \Delta t$.
As we will clarify in the next section, this definition implicitly assumes that the forcing term $\bm{b}$ does not affect the cantilever phase.
Using \eqnref{eq:propagator-average-frequency} and this definition, 
we recover a linear relationship between the cantilever phase and the change in the force-gradient by approximating $\theta$ to first order in $\kappa$ 
\begin{align}
\theta(t, t_0) & \approx 
    \omega_0 \, \Delta t
    + 
    \frac{\omega_0}{2} \int_{t_0}^{t} 
              \kappa(t') \: 
              d t',
\label{eq:propagator-phase}
\end{align}
where the approximation is justified because $\kappa \ll 1$.
We obtain the usual expression for the cantilever frequency in EFM by defining the cantilever's instantaneous frequency as the derivative of the cantilever phase:
\begin{equation}
f(t) =
\frac{1}{2\pi} \frac{d\theta}{dt} =
f_0-\frac{f_0}{4k_0}C'' V(t)^2,
\label{eq:KPFM-frequency}
\end{equation}
where, as usual, the voltage is $V(t) = V-\Phi$.
If $F(t)=0$, Eqs.~(\ref{eq:propagator-phase}) and (\ref{eq:KPFM-frequency}) hold even for arbitrarily fast changes to $\kappa(t)$.
In principle, then, there is no inherent limit to the time resolution that can be obtained from EFM measurements of the cantilever frequency or phase.
There are two potential complications, however.

First, it becomes very difficult to detect changes in the cantilever frequency \emph{directly}, by observing the cantilever's position over a short time interval, because the cantilever frequency measurement bandwidth must be smaller than the cantilever's resonance frequency \cite{Rihaczek1966mar,Boashash1992apr,Boashash1992apra}.
This seemingly fundamental bandwidth limitation can be surmounted by recording the phase shift as a function of a pulse delay \cite{Moore2009dec}, \latin{i.e.}\  \emph{indirectly}, as Dwyer and coworkers showed in the ``phasekick'' EFM experiment they introduced to measure fast, sub-cycle photocapacitance transients \cite{Dwyer2017jun}.
In the Ref.~\citenum{Dwyer2017jun} pk-EFM experiment (Fig.~\ref{fig:trEFM-and-pkEFM-schematic}),  a light pulse applied at time $t_0 = 0$ initiates charge generation in the sample.  
The capacitance derivative $C^{\prime\prime}$ is now time dependent, and the cantilever phase evolves in time according to \eqnref{eq:dk-kappa} and \eqnref{eq:propagator-phase}. 
At a time $t = t\st{p}$, the photo-induced advance of the cantilever phase is abruptly arrested stepping the tip voltage back to zero
\begin{equation}
V\st{t}(t) =
\begin{cases}
V & 
    \mathrm{for} \: \: \:
    t < \, t_{\text{p}} \\
0 & \mathrm{for} \: \: \:
    t \geq t_{\text{p}}.
\end{cases}
\label{Eq:Vt}
\end{equation}
The resulting cantilever phase is 
\begin{equation}
\theta(t\st{p})
 \approx
 \omega_0 t\st{p}
 - \frac{V^2}{2 m \omega_0} 
   \int_{0}^{t\st{p}} 
     C^{\prime\prime}(t^{\prime}) \,
   d t^{\prime}.  
\label{eq:pk-EFM-phase}
\end{equation}
In the Ref.~\citenum{Dwyer2017jun} experiment (Fig.~\ref{fig:trEFM-and-pkEFM-data}(d-f)), cantilever phase \latin{versus} time data were collected for a few milliseconds before and after the time window during which the light and voltage pulses were applied.
The phase shift $\theta(t\st{p})$ was obtained by extrapolating the ``before'' phase data to $t = 0$ and the ``after'' phase data to $t = t\st{p}$.
The pulse time $t\st{p}$ was stepped and this $\theta(t\st{p})$ measurement procedure was repeated at each $t\st{p}$.
Since $t\st{p}$, $\omega_0$, $m$, and $V$ are known, the full time-evolution of the sample's capacitance derivative $C^{\prime\prime}(t)$ could be inferred from the resulting $\theta(t\st{p})$ \latin{versus} $t\st{p}$ data.
In this way it is possible to track the evolution of photocapacitance on time scales much faster than a single cantilever cycle.

The second potential complication to measuring fast changes in cantilever frequency or phase is that abrupt changes to $F\st{ts}(t)$ cause additional changes in the cantilever's amplitude, frequency, and phase that we have so far neglected.
To address this problem, we first define the cantilever amplitude, phase and frequency in terms of the cantilever's position and momentum.

\section{Definition of amplitude and phase}
\label{sec:Defn-amp-phase}

EFM-based photocapacitance experiments record light-induced changes in the amplitude, phase, and frequency of the cantilever oscillation.
We define the cantilever amplitude and phase in terms of the cantilever position and momentum so we can relate the photocapacitive quantities $C'(t)$ and $C''(t)$ to the data.
We  show how abrupt changes in the tip-sample force $F\st{ts} = \frac{1}{2}C' (V-\Phi)^2$ affect the cantilever amplitude and phase.
The usual expression for the frequency shift in KPFM (Eq.~\ref{eq:Df-SKPM-const-V}) ignores these effects, which become important whenever the tip-sample force changes on a timescale similar to the cantilever period.

\begin{figure*}
\includegraphics[width=5.00in]{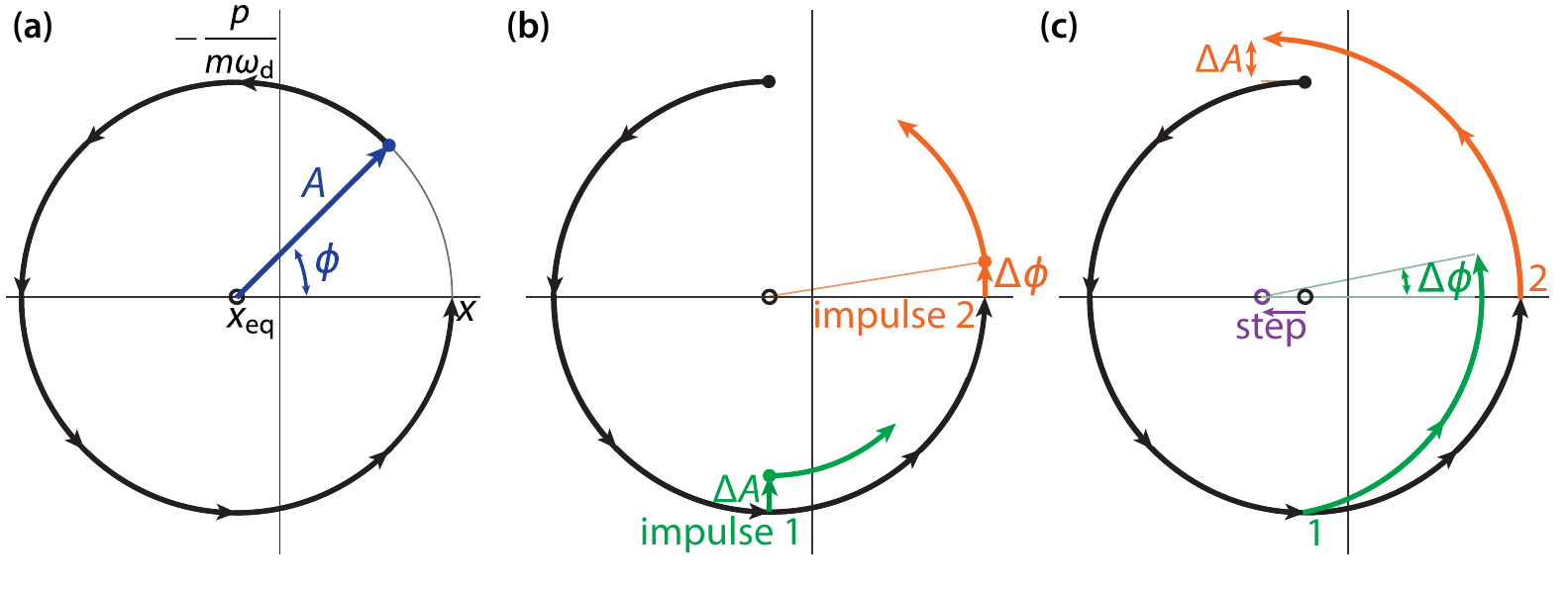}
\caption{Amplitude and phase representation of the cantilever state.
(a) The cantilever amplitude and phase are defined in terms of the cantilever position $x$ (horizontal axis), scaled momentum $-p/(m \omega\st{d})$ (vertical axis), and equilibrium position $x\st{eq}$ (open circle). 
The black line and arrows (every eighth of a period) show the normal evolution of the cantilever state in the absence of an abrupt change in tip-sample force.
(b) 
An impulsive force shifts the cantilever state along the vertical (momentum) axis.
An impulsive force applied at time 1 decreases the cantilever amplitude (green).
Applied at time 2, the same impulsive force instead advances the cantilever phase (orange).
(c) A step force shifts the equilibrium position $x\st{eq}$ (open circle).
Applying the step force at time 1 advances the cantilever phase (green), while applying the step force at time 2 increases the cantilever amplitude (orange).
}
\label{fig:xp-Aphi}
\end{figure*}

Figure~\ref{fig:xp-Aphi} provides a geometrical view of our definition of the cantilever amplitude and phase.
The horizontal axis shows the cantilever position $x$ and the vertical axis shows the scaled cantilever momentum $-p/(m \omega\st{d})$ (with $\omega\st{d}$ the drive frequency).
Each point on the graph is associated with a particular cantilever state $\bm{x} = (x \,\, p)^T$.
To define the cantilever amplitude and phase, however, we also need to know the equilibrium position that the cantilever state rotates about.
This equilibrium position---neglected in typical EFM experiments---is 
\begin{equation}
x\st{eq}(t) = \frac{F\st{ts}(t)}{k(t)},
\label{eq:x-eq-def}
\end{equation}
where $k(t)$ is the time-dependent cantilever spring constant.
We associate an amplitude and phase with each cantilever state $\bm{x}=(x \, \, p)^T$ using a complex number $z$:
\begin{equation}
z(t) = (x(t)-x\st{eq}(t)) - \frac{p(t)}{m \omega\st{d}} i,
\label{eq:z}
\end{equation}
where we assume a drive force of the form
\begin{equation}
F\st{dr}(t) = F\st{d} \cos(\omega\st{d} t + \phi\st{d}),
\label{eq:Fdrive}
\end{equation}
with $F\st{d}$ the drive amplitude, $\omega\st{d}$ the drive frequency, and $\phi\st{d}$ the drive phase.
In terms of the complex number $z$, the cantilever amplitude is
\begin{equation}
A = \abs{z},
\label{eq:A}
\end{equation}
and the absolute cantilever phase is
\begin{equation}
\phi\st{abs} = \arg{z}.
\label{eq:phi-abs}
\end{equation}
Fig.~\ref{fig:xp-Aphi}(a) shows this definition geometrically; the blue vector's length defines the amplitude and the angle with the $x$-axis defines the absolute phase.
With this definition, the ordinary evolution of the cantilever is 
$z(t) = z(t_0) e^{i \omega\st{d} (t-t_0)}$.
Graphically, the cantilever state vector has length $A$ and rotates around its equilibrium state $(x\st{eq}, 0)^{T}$ at the drive frequency $\omega\st{d}$.
To remove the effect of the ordinary evolution of the cantilever, we define the phase difference $\phi$ between the drive force and the cantilever displacement
\begin{equation}
\phi = \phi_{xp} = \arg \Big (z \, e^{-i (\omega\st{d} t + \phi\st{d})} \Big ).
\label{eq:phi}
\end{equation}
We use the subscript $\phi_{xp}$ to emphasize that this is the phase calculated from the cantilever position $x$ and momentum $p$.
With Eqs.~(\ref{eq:z}) and (\ref{eq:phi}), we can approximate the cantilever's phase using numerical simulations or analytic approximations of the cantilever position and momentum.

We use this definition of amplitude and phase to determine amplitude and phase shifts caused by abrupt forces.
We consider an experiment where the voltage, capacitance, tip-sample force, and tip-sample force gradient remain constant except for some short, abrupt change near $t=0$.
For times $t < 0$, the applied voltage $V$ induces a tip-sample force $F\st{ts}(t) = \frac{1}{2} C' V^2 \equiv F_0$ and a spring constant shift $\Delta k(t) = -\frac{1}{2} C'' V^2 \equiv \Delta k_0$.
The system is still a damped, driven harmonic oscillator but with a new spring constant
\begin{equation}
  k_1 = k_0 + \Delta k_0 
  \label{eq:k_1}
\end{equation}
and resonance frequency
\begin{equation}
\omega_1 = \omega_0 - \frac{\omega_0}{4 k_0} C'' V^2.
\label{eq:omega_1}
\end{equation}
We use a drive force with amplitude $F\st{d}$, frequency $\omega\st{d}$, and phase $\phi\st{d}$ (\eqnref{eq:Fdrive}).
The resulting cantilever state vector near $t = 0$ is
\begin{equation}
\bm{x}\st{ord} = \begin{pmatrix*}
x(t)\\
p(t)
\end{pmatrix*}
= \begin{pmatrix*}[r]
F_0/k_1 + A_0 \cos(\omega\st{d} t + \phi_0 )  \\
-A_0 m \omega\st{d} \sin(\omega\st{d} t + \phi_0)
\end{pmatrix*},
\label{eq:x-ord}
\end{equation}
where the equilibrium position is $x\st{eq} = F_0/k_1$ and the subscript $\bm{x}\st{ord}$ reflects that this is the cantilever's ordinary oscillation. 
The cantilever's amplitude $A_0 = \abs{\hat{\chi}(\omega\st{d})} F\st{d}$ and initial phase $\phi_0 = \phi\st{d} + \arg{\hat{\chi}(\omega\st{d})}$ depend on the Fourier transform of the oscillator's impulse response function
\begin{equation}
  \hat{\chi}(\omega) = \frac{1}{k_1}\left (1 - \frac{\omega^2}{\omega_1^2} +  \frac{2 j \gamma \omega}{\omega_1^2} \right )^{-1},
\label{eq:chi-omega1}
\end{equation}
where $\gamma$ is the linear damping parameter.
Eqs.~(\ref{eq:omega_1}--\ref{eq:chi-omega1}) describe the cantilever position, momentum, amplitude, and phase for a constant applied voltage.
To describe amplitude and phase shifts caused by abrupt forces, we consider adding an additional force $\Delta F\st{abrupt}$ at $t = 0$: $F\st{ts}(t>0) = F_0 + \Delta F\st{abrupt}(t)$.
Our model of the cantilever is linear so we can add the position and momentum change caused by this additional force to the ordinary, existing oscillation of the cantilever: $\bm{x}(t>0) = \bm{x}\st{ord} + \bm{x}\st{abrupt}$.
The change induced by the force is
\begin{equation}
\bm{x}\st{abrupt}(t) = \int_{0}^{t} \bm{U}(t, t') \begin{pmatrix} 0 \\ \Delta F\st{abrupt}(t') \end{pmatrix}.
\label{eq:x_abrupt}
\end{equation}

We consider two limits for the abrupt change in tip-sample force: an impulsive force and a step-like force.
For an impulsive force, the entire change in tip-sample force occurs over a very short time $t\st{impulse} \ll \omega_{1}^{-1}$.
The impulsive force changes the momentum of the cantilever by $\delta p = \int_0^{t\st{impulse}} \Delta F\st{abrupt}(t') dt'$.
In Figure~\ref{fig:xp-Aphi}(b), the impulse shifts the cantilever state along the vertical (momentum) axis.
An impulse delivered at time 1 shifts the cantilever amplitude (green), while the same impulse delivered at time 2 shifts the phase (orange).
After the impulse, the cantilever state continues rotating at the frequency $\omega\st{d}$.
If the change in momentum $\delta p$ is small ($\abs{\delta p / (m \omega\st{d})} \ll A_0$), the impulse shifts the cantilever amplitude and phase by
\begin{align}
\delta A &= -(m \omega\st{d})^{-1}\sin{\phi_0} \, \delta p 
\label{eq:dA-dp}
\,\,\, \text{ and}
\\
\delta \phi &= -(A_0 m \omega\st{d})^{-1} \cos{\phi_0} \, \delta p 
\label{eq:dphi-dp}
\end{align}
respectively.

\begin{figure*}
\includegraphics{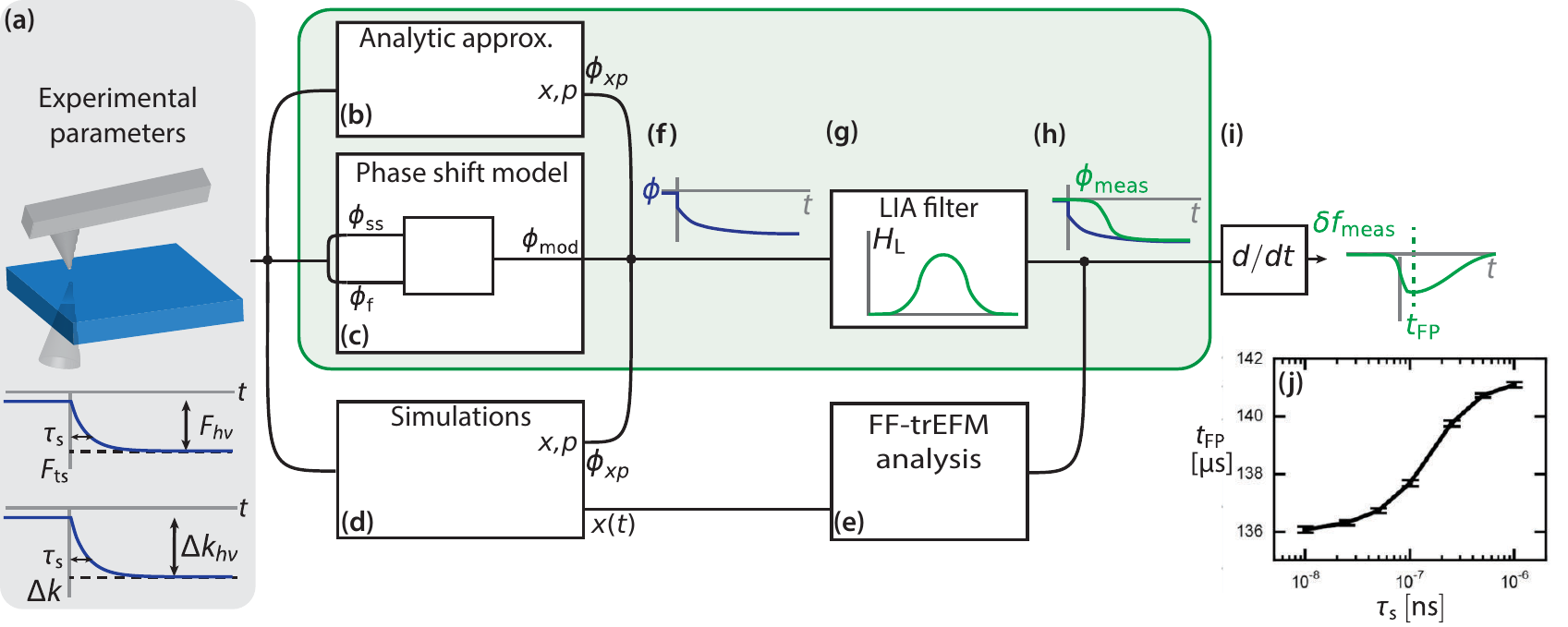}
\caption{Analysis of the feedback-free time-resolved electric force microscopy (FF-trEFM) experiment.
Part (j) reproduced with permission from Ref.~\citenum{Karatay2016may}.}
\label{fig:trEFM-01}
\end{figure*}

For an abrupt step-like force, the abrupt change in force is a constant: $\Delta F\st{abrupt}(t>0) = \Delta F\st{abrupt}$.
The abrupt change in force does not cause any instantaneous change in the cantilever state, but does induce an additional position oscillation $x\st{abrupt}(t) = \Delta F\st{abrupt} \cos(\omega_1 t) / k_1 $.
This additional oscillation is induced by the abrupt shift in the cantilever's equilibrium displacement by $\delta x\st{eq} = \Delta F\st{abrupt} / k_1$.
Figure~\ref{fig:xp-Aphi}(c) illustrates this result geometrically. The step change in tip-sample force abruptly shifts the cantilever equilibrium position $x\st{eq}$ (open purple circle denoted ``step'').
The actual cantilever state is not immediately affected by the abrupt change in force and the equilibrium position.
After the step, the cantilever state continues rotating at $\omega\st{d}$,\footnote{Here we assume that the drive frequency is at least approximately equal to the resonance frequency, $\omega\st{1} \approx \omega\st{d}$.}, but is rotating around the new equilibrium position (open purple circle).
Over the next part of the cantilever cycle, however, the effect of this change in equilibrium position becomes clear.
A step at time 1 shifts the cantilever phase (green), while the same step at time 2 shifts the cantilever amplitude (orange).
If the change in equilibrium position is small compared to the cantilever amplitude ($\abs{\delta x\st{eq}} \ll A_0$), we find that the step abruptly shifts the amplitude and phase by 
\begin{align}
\delta A& = -\cos{\phi_0} \, \delta x\st{eq}
\label{eq:dA-dx}
\,\,\, \text{ and}
\\
\delta \phi& = A_0^{-1}\sin{\phi_0} \, \delta x\st{eq} 
\label{eq:dphi-dx}
\end{align}
respectively.
According to our definition of amplitude and phase, $A$ and $\phi$ change abruptly as soon as the step force is applied.
However, the shift in phase cannot be readily observed in the cantilever position (horizontal axis) until perhaps $1/4$ of a cantilever cycle later (notice the time it takes for the difference between the black and green curve in Fig.~\ref{fig:xp-Aphi}(c) to develop).
The cantilever amplitude and phase are not well-determined during the short period of time when the the tip-sample force is changing abruptly.
As discussed in Ref.~\citenum{Dwyer2017jun}, changes in tip-sample forces occurring on a timescale short compared to the cantilever period should be detected using a measurement that exploits a nonlinearity to generate a signal that can be measured at low frequency.

For both impulsive and step-like changes to the tip-sample force, the resulting change in the cantilever's position and momentum can be determined by integrating \eqnref{eq:x_abrupt}.
After the end of the abrupt changes in tip-sample force, the two components of the cantilever state vector can be added back together and propagated as usual.
In each case, the same shift in equilibrium position or momentum can cause an amplitude shift or a phase shift depending on the cantilever's initial phase when the abrupt force occurs, as illustrated by the geometric depiction in Fig.~\ref{fig:xp-Aphi}(b) and (c).
The change in position or momentum affects the cantilever phase more when the cantilever's initial amplitude $A_0$ is smaller.
The results of the previous two sections allow us to analyze experiments involving abrupt changes in the tip-sample force and force gradient. 
In Ref.~\citenum{Dwyer2017jun}, we analyzed the phasekick electric force microscopy experiment (pk-EFM), which was developed to measure light-induced changes in capacitance with sub-cycle time resolution.
In the next section, we analyze the alternative technique, FF-trEFM.

\section{FF-\lowercase{tr}EFM time resolution}
\label{Sec:FF-trEFM}

Feedback-free time-resolved electric force microscopy (FF-trEFM) \cite{Giridharagopal2012jan,Cox2015aug,Karatay2016may} is a variant of tr-EFM designed to resolve photocapacitance dynamics with better time resolution.
Ordinary tr-EFM measurements directly fit the cantilever-frequency-shift-\latin{versus}-time data to extract the sample's photocapacitance risetime $\tau\st{s}$.
In FF-trEFM, the cantilever is driven at a fixed frequency $\omega\st{d}$ with a fixed tip voltage $V$.
The light is turned on at a specific point in the cantilever cycle and the cantilever oscillation data is signal-averaged and demodulated to obtain the cantilever's instantaneous frequency shift $\delta f$ versus time.
The time-to-first-frequency-shift peak $t\st{FP}$ is calculated from $\delta f(t)$ (Fig.~\ref{fig:overview}(c)).
To calibrate the measurement, voltage pulses with different rise times $\tau\st{v}$ are applied to the sample and $t\st{FP}$ is measured versus $\tau\st{v}$.
The sample photocapacitance risetime $\tau\st{s}$ is estimated using the $t\st{FP}$ versus $\tau\st{v}$ calibration curve.
Ginger and co-workers have shown that sub-cycle time resolution can be obtained with this technique.
Through numerical simulations, they demonstrated that the effect of the cantilever tip-sample force $F\st{ts} \propto C'$ gives rise to the sub-cycle time resolution \cite{Karatay2016may}.

In this section, we apply our Magnus expansion approximation for cantilever dynamics to the FF-trEFM experiment.
We show that the FF-trEFM experiment is only sensitive to the total magnitude of the force-induced phase shift at short times.
To extract a specific time constant in the limit that $\tau\st{s} \ll \omega_0^{-1}$, an assumption must be made about the magnitude of the abrupt change in the tip-sample force.
This result demonstrates how our approach reveals the hidden assumptions implicit in commonly used models of EFM experiments.

Figure~\ref{fig:trEFM-01} illustrates our analysis of the FF-trEFM experiment.
To explain the origin of the sub-cycle time resolution in FF-trEFM, we need to connect the experimental and sample parameters (Fig.~\ref{fig:trEFM-01}(a)) to the measured frequency-shift-\latin{versus}-time data that is used to calculate the time-to-first-frequency-shift-peak $t\st{FP}$.
We start from the description of EFM derived in Sec.~\ref{Sec:Magnus}, assuming the tip-sample force $F\st{ts}$ and force gradient $\Delta k$ both evolve with the same photocapacitance risetime $\tau\st{s}$ after the light turns on (Fig.~\ref{fig:trEFM-01}(a)).
We use the tip-sample force and force gradient to determine the cantilever's phase $\phi$ in multiple ways (Fig~\ref{fig:trEFM-01}(b--d)).
To isolate the effect of the sample parameters on the phase and frequency, we model the \emph{measured} cantilever phase as the convolution of the cantilever's actual phase  and a demodulation or lock-in amplifier low-pass filter (Fig.~\ref{fig:trEFM-01}(f--h)).
The measured cantilever phase shift is
\begin{equation}
\phi\st{meas}(t) = [H\st{L} \ast \phi](t),
\label{eq:phi-meas}
\end{equation}
where $H\st{L}$ is the lock-in amplifier or demodulation filter impulse response function, $\ast$ denotes convolution in the time domain, and $\phi$ is the actual phase difference between the tip displacement and drive force.
The measured cantilever frequency shift is the derivative of the measured phase
\begin{equation}
\delta f \st{meas}(t) = \frac{1}{2\pi} \d{\phi\st{meas}}{t}.
\label{eq:df-meas}
\end{equation}
At the time of the first frequency shift peak ($t=t\st{FP}$), the derivative of the measured frequency shift is equal to zero (Fig.~\ref{fig:trEFM-01}(i)).
The specific value of $t\st{FP}$ is sensitive to the choice of lock-in amplifier filter $H\st{L}$.
Once a particular $H\st{L}$ is chosen, any differences in $t\st{FP}$ are related to differences in the cantilever's actual phase $\phi(t)$.
To avoid artifacts in $\phi\st{meas}$ and $\delta f\st{meas}$ related to filter ringing, Karatay and co-workers used a filter function that was strictly positive \cite{Karatay2016may}.
As shown in Fig.~\ref{fig:trEFM-01}(j), they observed a monotonic, nonlinear relationship between $t\st{FP}$ and $\tau\st{s}$ for risetimes faster than the cantilever period of $\SI{2}{\us}$ under carefully chosen experimental conditions.
We connect the observed $t\st{FP}$ to $\phi(t)$ and therefore to experimental parameters using the model of the cantilever dynamics developed in Sections~III and IV.

In the following calculation, we will verify the relationship between experimental parameters and $\phi$ by estimating the phase in multiple ways.
In particular, we estimate the phase from simulations of the cantilever position and momentum (Fig.~\ref{fig:trEFM-01}(d)).
We separately analyze the simulated cantilever position data using the Ginger group's analysis code \cite{Karatay2015}.
We calculate the same $t\st{FP}$ with both approaches, which connects our new analysis (shaded green region of Fig.~\ref{fig:trEFM-01}) to that used by Ginger and co-workers.

\subsection{Analytic treatment of FF-trEFM}

We use the description of the cantilever amplitude and phase developed in the previous section to determine the cantilever phase during an FF-trEFM experiment.
The applied voltage is $V$, and the drive force has amplitude $F\st{d}$, frequency $\omega\st{d}$, and phase $\phi\st{d}$ (\eqnref{eq:Fdrive}). 
At $t = 0$, the sample is illuminated, inducing a change in the tip-sample capacitance and its derivatives, which we assume has the form
\begin{equation}
C'(t) = \begin{cases}
C' & t < 0 \\
C' + \Delta C'_{h\nu} ( 1 - e^{-t/\tau\st{s}}) & t \geq 0
\end{cases},
\label{eq:Czt-FFtrEFM}
\end{equation}
where $\Delta C'_{h\nu}$ is the light-induced change in the tip-sample capacitance at long times and $\tau\st{s}$ is the sample's photocapacitance risetime.
Similarly, we assume the second derivative of the tip-sample capacitance is
\begin{equation}
C''(t) = \begin{cases}
C'' & t < 0 \\
C'' + \Delta C''_{hv} ( 1 - e^{-t/\tau\st{s}}) & t \geq 0
\end{cases},
\label{eq:Czzt-FFtrEFM}
\end{equation}
where $\Delta C''_{h\nu}$ is the light-induced change in the second derivative of the capacitance.

The goal of the experiment is to infer $\tau\st{s}$ from the first frequency-shift-peak time $t\st{FP}$.
We analyze the experiment using the Magnus-expansion approximation (Sec.~\ref{Sec:Magnus}).
The exponential rise in $C'(t)$ causes an exponential rise in the tip-sample force
\begin{equation}
F\st{ts}(t) = \begin{cases}
F_0 & t < 0 \\
F_0 + F_{h\nu} ( 1 - e^{-t/\tau\st{s}}) & t \geq 0
\end{cases},
\label{eq:Fts-exp}
\end{equation}
where the initial force is $F_0 = \frac{1}{2} C' V^2$ and the light-induced change in force is $F_{h\nu} = \frac{1}{2} \Delta C'_{h\nu} V^2$.
The exponential rise in $C''(t)$ causes an exponential rise in the tip-sample force gradient
\begin{equation}
\Delta k(t) = \begin{cases}
\Delta k_0 & t < 0 \\
\Delta k_0 + \Delta k_{h\nu} ( 1 - e^{-t/\tau\st{s}}) & t \geq 0
\end{cases},
\label{eq:dk-exp}
\end{equation}
where the initial spring constant shift is $\Delta k_0 = \frac{1}{2} C'' V^2$ and the light-induced change in the spring constant shift is $\Delta k_{h\nu} = \frac{1}{2} \Delta C''_{h\nu} V^2$.

Our goal is to understand and explain the case where the photocapacitance risetime is faster than the cantilever period. In this case, we expect that the step-like change in tip-sample force to cause an abrupt change in the cantilever amplitude and phase (Fig.~\ref{fig:xp-Aphi} and Eqs.~(\ref{eq:dA-dx}--\ref{eq:dphi-dx})).
To focus on the effect of the tip-sample force $F\st{ts}$, we first assume $\Delta C''_{h\nu} =0$ so that $\Delta k(t) = \Delta k_0$.
We are interested in times much shorter than the cantilever ringdown time ($t \ll \gamma^{-1}$) so we neglect cantilever dissipation and the drive force by setting $\gamma = 0$ and $F\st{dr}(t)=0$.
The Magnus expansion approximation for the cantilever state (\eqnref{eq:propagator-evolve-state-plus-forcing-term}) is 
\begin{equation}
\bm{x}(t)
= \bm{\widetilde{U}}(t) \,
\begin{pmatrix}
x_0 \\
p_0
\end{pmatrix}
+ \int_{0}^{t}
\bm{\widetilde{U}}(t-t') \,
\begin{pmatrix}
0 \\
F\st{ts}(t')
\end{pmatrix} 
dt',
\label{eq:abrupt-force-response}
\end{equation}
where
\begin{equation}
\bm{\widetilde{U}}(t)
=
\begin{pmatrix*}[c]
\cos{\left ( {\omega_1}
     t \right )} \hphantom{m} & 
\dfrac{1}{m {\omega_1}}\sin{\left( {\omega_1} 
      t \right)} \\
- m {\omega_1}
\sin{\left( {\omega_1}
     t \right)} \hphantom{m} &
\cos{\left( {\omega_1} 
     t \right)}
\end{pmatrix*}
\label{eq:abrupt-propagator}
\end{equation}
is the propagator of a simple harmonic oscillator with a shifted spring constant $k_1$ and resonance frequency $\omega_1$ given by Eqs.~(\ref{eq:k_1}) and (\ref{eq:omega_1}) respectively.
The integral in \eqnref{eq:abrupt-force-response} can be evaluated in closed form, giving
\begin{multline}
x(t) = \left (x_0 - \frac{F_0}{k_1} \right ) \cos (\omega_1 t) + \frac{p_0}{m \omega_1} \sin (\omega_1 t)  \\
 + \frac{F_0}{k_1} + \frac{F_{h\nu}}{k_1} \left ( 1 - \frac{\omega_1^2 \tau\st{s}^2}{1 + \omega_1^2 \tau\st{s}^2} e^{-t/\tau\st{s}} \right ) \\
 \underbrace{-\frac{F_{h\nu}}{k_1 \left(1+\omega_1^2 \tau\st{s}^2\right)} \Big (
\cos (\omega_1 t)
+ \omega_1 \tau\st{s} \sin (\omega_1 t) 
\Big )}_{x\st{osc}}
\label{eq:x-f-osc}
\end{multline}
for the cantilever position.
The first line of \eqnref{eq:x-f-osc} is the unperturbed continuation of the cantilever's ordinary oscillation $x\st{ord}(t)$ (\eqnref{eq:x-ord}).
The second line describes the change in the cantilever's time-dependent equilibrium position $x\st{eq}$ (\eqnref{eq:x-eq-def}).
The final, underbraced line ($x\st{osc}$) is the persistent contribution of the step change in $F\st{ts}(t)$ to the cantilever oscillation, which, depending on the phase of the cantilever oscillation at $t=0$, manifests as an abrupt amplitude or phase shift.

\begin{figure}
\includegraphics[width=3.25in]{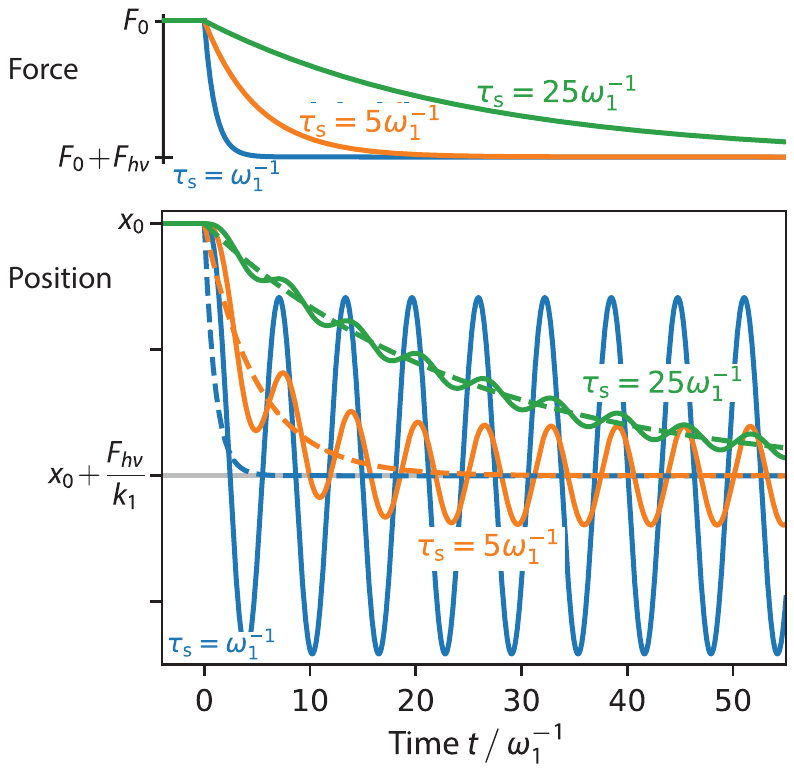}
\caption{Cantilever response to step changes in force with different risetimes $\tau\st{s}$.
The (a) tip-sample force $F\st{ts}$ and (b) cantilever position $x(t)$ \latin{versus} time $t$ (in units of $\omega_1^{-1}$).
The solid lines in (b) show the position calculated using \eqnref{eq:x-f-osc}.
For reference, the dashed lines in (b) show $x = F\st{ts}(t)/k_1$, the cantilever position for a cantilever with a much higher resonance frequency such that $F\st{ts}$ changes slowly relative to the cantilever period.
}
\label{fig:01-force-FFtrEFM}
\end{figure}

Figure~\ref{fig:01-force-FFtrEFM} illustrates the effect of the abrupt change in tip-sample force on the cantilever oscillation for various time constants $\tau\st{s}$.
We focus on the effect of the abrupt, step-like change in tip-sample force by starting the cantilever at rest at its equilibrium position
\footnote{The initial position is $x_0 = F_0/k_1$ and initial momentum is $p_0 = 0$.}
so that the first line of Eq.~\ref{eq:x-f-osc} equals zero.
The solid lines in Fig.~\ref{fig:01-force-FFtrEFM} plot the cantilever position versus time for the different risetime forces.
The cantilever position contains both an exponential component (the dashed line) and an oscillatory component.
The oscillatory component is largest in magnitude for the shortest rise time forces and approaches zero for $\tau\st{s} \gg \omega_1^{-1}$.
From \eqnref{eq:x-f-osc}, the amplitude of the induced oscillation is $F_{h\nu} / \big( k_1 (1+\omega_1^2 \tau\st{s}^2) \big )$.
When the tip-sample force changes abruptly, the ordinary KPFM frequency shift given by \eqnref{eq:Df-SKPM-const-V} is incomplete; the cantilever oscillation is better described using the cantilever phase, which is advanced by both the usual frequency shift given by \eqnref{eq:Df-SKPM-const-V}, as well as the abrupt effects described in X, Y.

The two forces on the cantilever, the drive force and the tip-sample capacitance force, affect the cantilever's phase very differently.
The drive force determines the cantilever's oscillation frequency.
Together, the properties of the drive force and the propagator determine the cantilever's amplitude and phase difference relative to the drive force.
In contrast, $F\st{ts}$ only determines the equilibrium displacement about which the cantilever oscillates unless its contains significant energy at the cantilever resonance frequency.

To apply these results of this section to ion-conductance experiments, the  exponential risetime change in capacitance could be replaced with a stretched exponential risetime change in capacitance by replacing $e^{-t/\tau\st{s}}$ with $e^{-(t/\tau\st{s})^\beta}$ in Eqs.~(\ref{eq:Czt-FFtrEFM}--\ref{eq:dk-exp}).
In this case, the integral in Eq.~(\ref{eq:abrupt-force-response}) cannot be evaluated in closed form.
The magnitude and phase of the induced oscillation at the cantilever frequency could be determined by numerical integration or using the Laplace transform of the stretched exponential \cite{Lindsey1980oct}.

\subsection{Approximate phase shift model}
\label{Sec:approximate-phase-mod}

To gain insight into the dynamics of the cantilever phase, we develop an approximate model to describe small cantilever phase shifts (Fig.~\ref{fig:trEFM-01}(c)).
When the light is turned on, changes in capacitance affect the phase difference between the drive force and the cantilever through (1) changes in the tip-sample force gradient and (2) abrupt changes in the tip-sample force.

The changes in the force gradient shift the cantilever's natural resonance frequency $\omega(t)$, which results in a phase shift.
For $\kappa = \Delta k / k_0 \ll 1$, the cantilever's resonance frequency is
\begin{equation}
\omega(t) = \omega_0 (1 +  \Delta k(t)/(2 k_0)),
\label{eq:omega-t}
\end{equation}
where we differentiate  Eq.~(\ref{eq:propagator-phase}) to obtain Eq.~(\ref{eq:omega-t}).
At steady state, the phase difference between the cantilever and the drive may be computed from the Fourier transform of the oscillator impulse response function.
We find
\begin{equation}
\phi\st{ss}(t) = \arg \Bigg( \left[ 1-\frac{\omega\st{d}^2}{\omega(t)^2} + \frac{i \omega\st{d}}{Q \omega(t)} \right]^{-1} \Bigg).
\label{eq:phi-ss}
\end{equation}
For small phase shifts, the cantilever response to changes in $\phi\st{ss}$ is first order with a characteristic frequency equal to the linear damping parameter $\gamma = \omega_0/(2 Q)$
\begin{equation}
\dot{\phi}\st{fg} = - \gamma \phi\st{fg} + \gamma \phi\st{ss}(t),
\label{eq:fg}
\end{equation}
where we use the subscript ``fg'' for force gradient.
The drive force and force gradient induce a slow evolution of $\phi\st{fg}$.

\begin{figure*}
\includegraphics{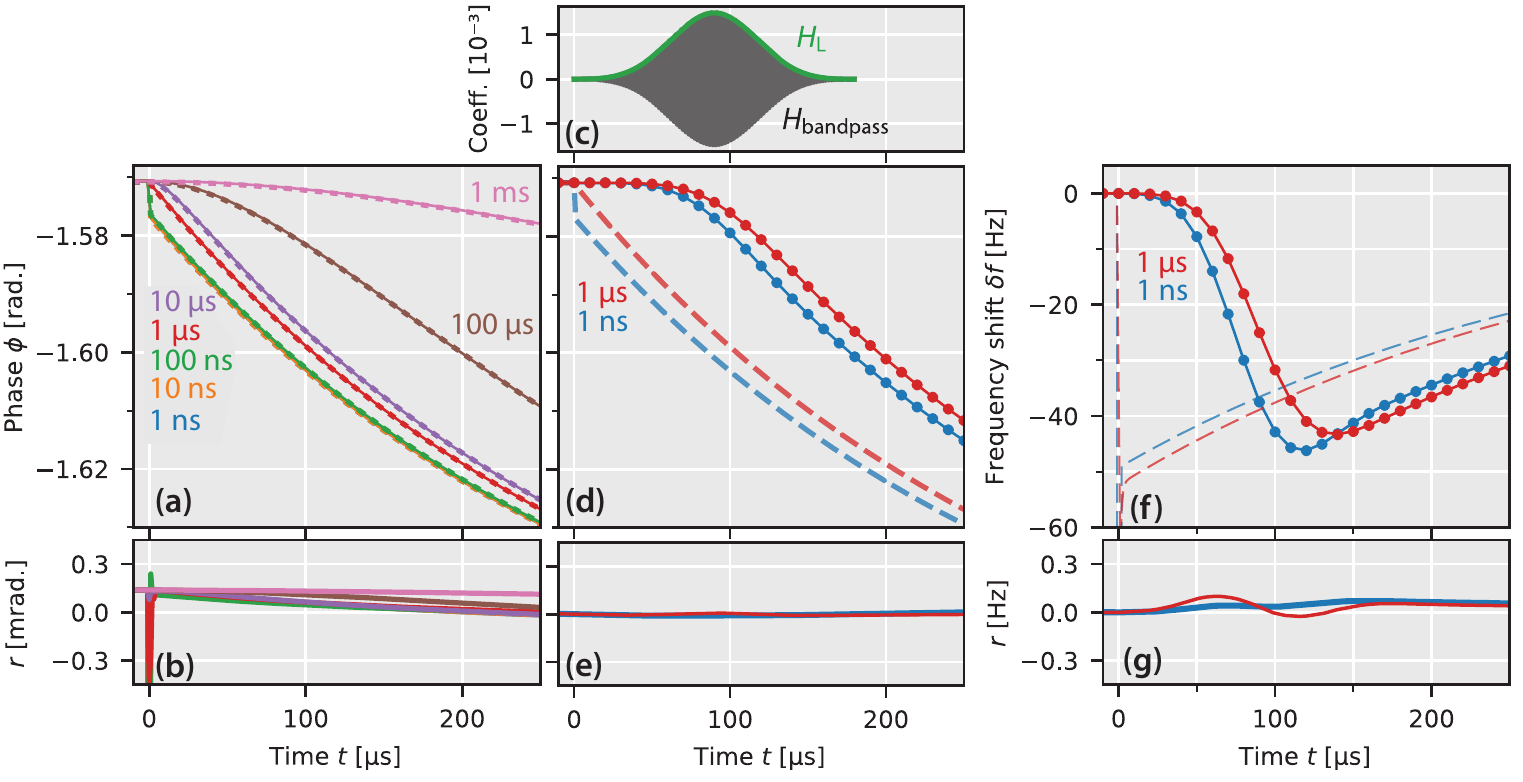}
\caption{
Actual and measured cantilever phase and frequency calculated using models and simulations.
(a) Relative phase $\phi$ between the cantilever position and the drive force for sample photocapacitance risetime constants $\tau\st{s}$ from \SI{1}{\ns} to \SI{1}{\ms}.
The phase was calculated from simulations of cantilever position and momentum ($\phi_{xp}$, solid lines) and the approximate small-phase-shift model of \eqnref{eq:phi-fg-f} ($\phi\st{mod}$, dashed lines).
(b) Phase difference between the two models $r = \phi_{xp} - \phi\st{mod}$.
(c) The lock-in amplifier filter (green line) used to determine $\phi\st{meas}$ from $\phi$ and the bandpass filter used to determine $\phi\st{meas}$ from simulated cantilever position data.
(d) The actual phase shift $\phi\st{mod}$ (dashed lines) and the measured phase shift calculated from $\phi\st{mod}$ (solid lines) and from simulated position versus time data using the Ginger group's analysis package FFTA with bandpass filter $H\st{bandpass}$ (circles).
(e) Phase difference between the two calculated, measured phases $r = \phi^\mathrm{mod}\st{meas} - \phi^\mathrm{FFTA}\st{meas}$.
(f)
Actual frequency shift calculated from $\phi\st{mod}$ (dashed lines) and the corresponding measured frequency shift calculated from $\phi^\mathrm{mod}\st{meas}$ (solid lines) and $\phi^\mathrm{FFTA}\st{meas}$ (circles).
(g)
Frequency shift difference $r = \delta f^\mathrm{mod}\st{meas} - \delta f^\mathrm{FFTA}\st{meas}$.
For simulated data (circles in (d) and (f)), only every thousandth point (every \SI{10}{\us}) is shown for clarity. 
}
\label{fig:trEFM-02}
\end{figure*}

The change in tip-sample force affects the cantilever phase differently.
Abrupt changes in the tip-sample force $F\st{ts}$ induce an additional oscillation at the cantilever resonance frequency (Fig.~\ref{fig:01-force-FFtrEFM}).
For the FF-trEFM experiment, the existing oscillation near $t=0$ is $x(t) = -A_0 \sin(\omega\st{d} t)$.
so the cosine term from Eq.~(\ref{eq:x-f-osc}) causes an abrupt phase shift
\begin{equation}
\Delta \phi\st{f} = \frac{F_{h\nu} / k_1}{A_0} \frac{1}{1 + \omega_{1}^2 \tau\st{s}^2},
\label{eq:delta-phi-f}
\end{equation}
where $A_0$ is the cantilever zero-to-peak amplitude near $t=0$, and $\Delta \phi\st{f}$ is in units of radians.
For times $t \leq \tau\st{s}$, the phase shift oscillates and approaches $\Delta \phi\st{f}$.
For the sake of our model, we assume
\begin{equation}
\phi\st{f}(t \geq 0) = \Delta \phi\st{f} ( 1 - e^{-t/\tau\st{s}}).
\label{eq:delta-phi-f-t}
\end{equation}

With the force contribution to the cantilever phase accounted for, we need to correct Eq.~(\ref{eq:fg}) to take into account $\phi\st{f}$.
The total cantilever phase is the sum of the force-gradient phase $\phi\st{fg}$ and $\phi\st{f}$:
\begin{equation}
\phi = \phi\st{fg} + \phi\st{f}(t).
\label{eq:phi-fg-f}
\end{equation}
We describe the combined effects of the force and force-gradient terms with the differential equation
\begin{align}
\dot{\phi}\st{fg}& = - \gamma \big (\phi\st{fg} + \phi\st{f}(t) \big)  + \gamma \phi\st{ss}(t).
\label{eq:dotphi}
\end{align}
If the phase $\phi = \phi\st{fg} + \phi\st{f}$ is equal to the steady state phase $\phi\st{ss}$, the derivative $\dot{\phi}\st{fg} = 0$ and the normal oscillator dynamics do not change the cantilever phase.
Together, Eqs.~(\ref{eq:phi-fg-f}) and (\ref{eq:dotphi}) describe a state space model with two inputs $\phi\st{ss}(t)$ and $\phi\st{f}(t)$, one state variable $\phi\st{fg}$, and one output $\phi$.
With this model, we can write closed-form expressions for the cantilever phase when exponential risetime inputs are applied to $\phi\st{ss}$ and $\phi\st{f}$.
 With either simulations of the cantilever position and momentum, or the approximate phase shift model, we can write the cantilever's \emph{actual} phase $\phi$.

\subsection{Simulations}

To verify the phase model developed above, we simulated cantilever dynamics for a cantilever similar to that performed by Karatay and co-workers in their demonstration of \SI{10}{\ns} time resolution \cite{Karatay2016may}.
We used a cantilever frequency at $t=0$ equal to $\omega_1 = 2 \pi \times \SI{526315}{\Hz}$.
The cantilever spring constant and quality factor were $k_1 = \SI{72.7}{\N\per\m}$ and $Q = \num{499}$ respectively.
We set the drive frequency $\omega\st{d} = \omega_1$.
We used a drive amplitude $F\st{d} = k_1 \times \SI{10}{\nm} / Q$ and a drive phase $\phi\st{d} = \pi$ for maximum time resolution.\footnote{The initial phase difference between the cantilever and drive is $-\pi/2$ since $\omega\st{d} = \omega_{1}$. Setting the drive phase $\phi\st{d} = \pi$ ensures the cantilever oscillation near $t=0$ is $x = A_0 \cos\left(\omega\st{d} + \pi - \pi/2\right) = -A_0 \sin(\omega\st{d} t)$, the phase of maximum time resolution in Refs.~\citenum{Giridharagopal2012jan} and \citenum{Karatay2016may}.}
The cantilever's simulated zero-to-peak amplitude at $t = 0$ was $A_0 = \SI{10}{\nm}$.
The light-induced change in spring constant was $\Delta k_{h\nu} = - 2 k_1 \times 10^{-4}$, corresponding to a cantilever frequency shift of $\Delta \omega_{h\nu} = 2 \pi \times \SI{52.6}{\Hz}$.
The light-induced change in the tip-sample force was $F_{h\nu} = -k_1 \times \SI{0.06}{\nm}$, inducing a $\SI{0.06}{\nm}$ shift in the cantilever's equilibrium displacement $x\st{eq}$.

\subsection{Results}

Figure~\ref{fig:trEFM-02} demonstrates the close agreement between the different models for the cantilever phase shift illustrated in Fig.~\ref{fig:trEFM-01}.
In Fig.~\ref{fig:trEFM-02}(a), we show the results of simulations for a series of sample photocapacitance risetimes $\tau\st{s}$ from $\SI{1}{\ns}$ to $\SI{1}{\ms}$.
From the simulated cantilever position and momentum, we calculated the cantilever phase $\phi_{xp}$ using Eq.~(\ref{eq:phi}).
In Fig.~\ref{fig:trEFM-02}(a), we plot $\phi_{xp}$ convolved with a rectangular filter with width $T = 2 \pi / \omega_0$ to remove phase oscillations at multiples of the cantilever frequency (solid lines).
We also plot the modeled phase $\phi\st{mod}$ (dashed lines), which was calculated using the approximate phase model of Sec.~\ref{Sec:approximate-phase-mod} (Eqs.~(\ref{eq:phi-ss})--(\ref{eq:dotphi})).
The simulated phase $\phi_{xp}$ agrees closely with the phase predicted by the analytic model $\phi\st{mod}$.
Figure~\ref{fig:trEFM-02}(b) shows that the phase difference $r = \phi_{xp} - \phi\st{mod}$ is small and approaches zero at long times.
Both the analytical model and simulations indicate that the cantilever phase versus time is identical for any photocapacitance risetime $\tau\st{s} \leq \SI{100}{\ns}$.
This sets the first limit on the possible time resolution of FF-trEFM.
Figure~\ref{fig:trEFM-02}(a,b) demonstrates good agreement between our different models of the cantilever's actual phase (Fig.~\ref{fig:trEFM-01}(b--d)).

Next we determine the \emph{measured} phase shift $\phi\st{meas}$.
First we calculate $\phi\st{meas}$ using $\phi$ from Fig.~\ref{fig:trEFM-02}(a) and the convolution model illustrated in Fig.~\ref{fig:trEFM-01}(f--h) (Eq.~(\ref{eq:phi-meas})).
To demonstrate the agreement between this description of the phase and the phase calculated from the FF-trEFM workup, we use the same simulation data used to calculate $\phi_{xp}$ to perform the FF-trEFM analysis of Ginger and co-workers using their publicly available package  \cite{Karatay2015,Karatay2016may}.
The bandpass filter applied to the $x(t)$ data in the FF-trEFM analysis serves the same role as the low-pass filter $H\st{L}$ in our analysis.
For the FF-trEFM data analysis protocol of Ref.~\citenum{Karatay2016may}, we use a Parzen window bandpass filter that passes frequencies between $f_0 - b$ and $f_0 + b$ (Fig.~\ref{fig:trEFM-01}(e)).
We use the analogous Parzen window low-pass filter with cutoff frequency $b$ in our analyses.
Figure \ref{fig:trEFM-02}(c) shows the two filters, with $b = \SI{5.1}{\kHz}$.
Figure \ref{fig:trEFM-02}(d) shows that the measured phases calculated using the model (solid lines) and the FFTA analysis (dot-dashed lines) agree closely.
For comparison, the input to the low-pass filter $\phi\st{mod}$ (dashed lines) is also shown.
The low-pass filter blurs and delays the phase.
Figure~\ref{fig:trEFM-02}(e) shows the maximum difference between $\phi\st{meas}^\mathrm{mod}$ and $\phi\st{meas}^\mathrm{FFTA}$ is \SI{8}{\micro\radian}.
Figure~\ref{fig:trEFM-02}(f,g) shows that the corresponding measured frequency shifts and time to first frequency shift peaks agree closely as well.
The maximum frequency difference is \SI{0.1}{\Hz}.

The data of Fig.~\ref{fig:trEFM-02} demonstrate that the new models we introduced to describe the measured phase in the FF-trEFM experiment agree closely with the measured phase as calculated by Ginger and co-workers.
We examine the dependence of $\phi$ on experimental parameters to better understand how the experimental parameters affect the measured phase and time-to-first frequency-shift peak.

\begin{figure}
\includegraphics{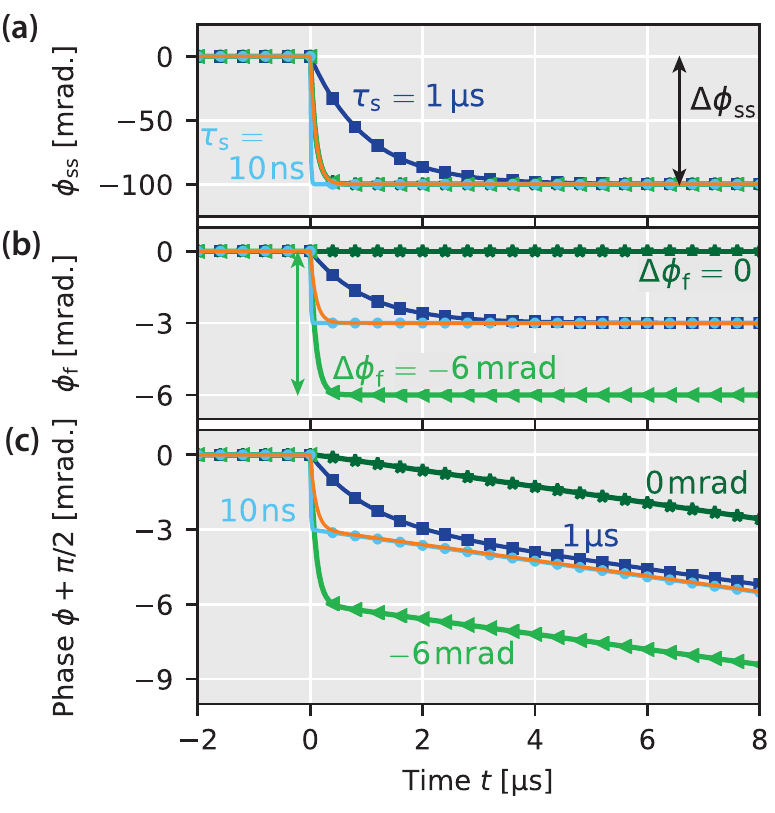}
\caption{Dependence of $\phi$ on photocapacitance risetime and force-induced phase shift.
(a) The steady-state photocapacitance $\phi\st{ss}$ for different photocapacitance risetimes (\SI{10}{\ns}, light blue circles; \SI{100}{\ns}, orange line; \SI{1}{\us}, dark blue squares).
(b) The force-induced phase shift for different photocapacitance risetimes (blue) and different magnitudes of $\Delta\phi\st{f}$ (\SI{-6}{\milli\radian}, light green triangles; \SI{-3}{\milli\radian}, orange line; \SI{0}{\milli\radian}, dark green stars).
(c) The phase determined from the inputs in (a) and (b), offset by the initial phase shift $\pi/2$.
}
\label{fig:trEFM-03}
\end{figure}

The analytic model of Eqs.~(\ref{eq:phi-fg-f}) and (\ref{eq:dotphi}) gives a closed-form expression for the cantilever phase during a FF-trEFM experiment.
In the analytic model, the cantilever's actual phase is
\begin{multline}
\phi(t \geq 0) = \phi_0 + \Delta \phi\st{ss} - \Delta \phi\st{f} e^{-t/\tau\st{s}}
\\
+ \frac{\Delta \phi \st{ss} - \Delta \phi \st{f}}{1 - \gamma \tau\st{s}} \big (
\gamma \tau\st{s} e^{-t/\tau\st{s}} - e^{-\gamma t}
\big ),
\label{eq:phi-mod-closed-form}
\end{multline}
where $\phi_0$ is the phase difference between the cantilever and drive force at $t = 0$, $\Delta \phi\st{ss}$ is the steady-state phase shift $\Delta \phi\st{ss} = \phi\st{ss}(\infty) - \phi\st{ss}(0)$ (Eq.~(\ref{eq:phi-ss})) and $\Delta \phi\st{f}$ is the total phase shift induced by the abrupt change in the tip-sample force (Eq.~(\ref{eq:delta-phi-f})).

\begin{figure*}
\includegraphics{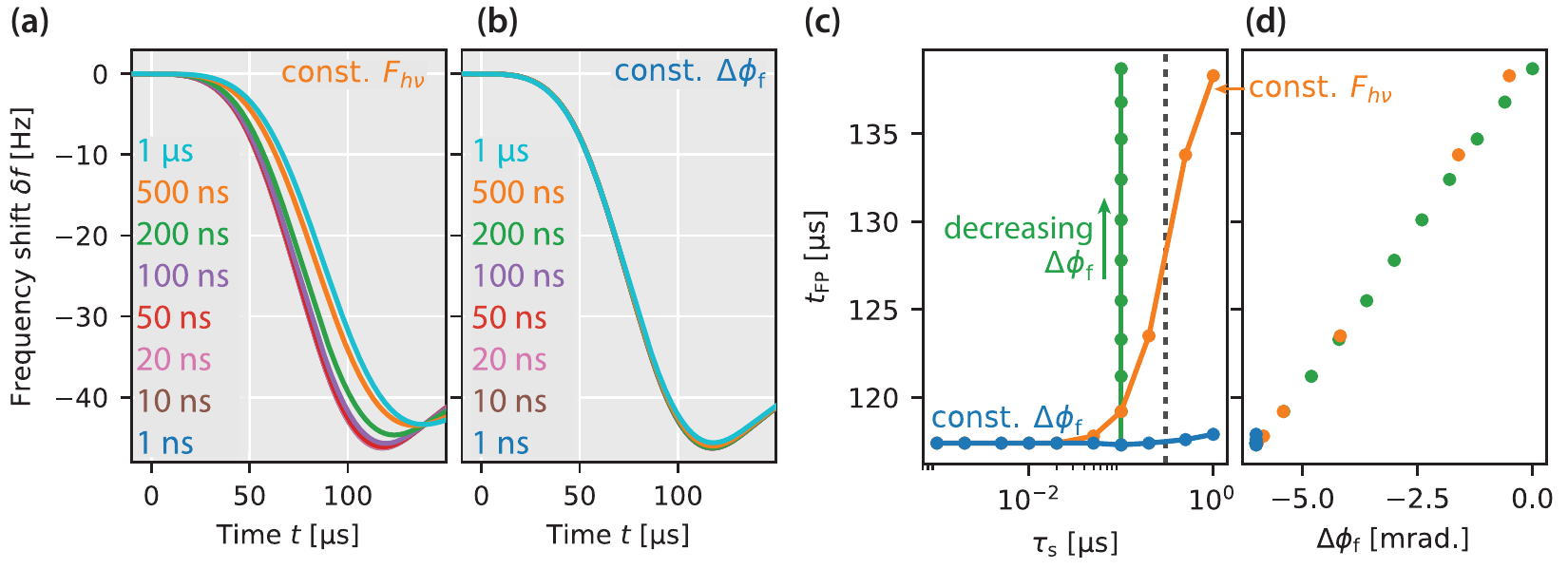}
\caption{(a) Demodulated frequency shift \latin{versus} pulse time for simulated photocapacitance dynamics with risetime $\tau\st{s}=\SI{1}{\ns}$ to $\tau\st{s} =\SI{1}{\us}$, keeping the light-induced tip-sample force $F_{h\nu} = -k_0 \times \SI{0.06}{\nm}$.
(b) Demodulated frequency shift \latin{versus} pulse time for simulated photocapacitance dynamics with risetime $\tau\st{s}=\SI{1}{\ns}$ to $\tau\st{s} =\SI{1}{\us}$ keeping the total phase shift induced by the change in  tip-sample force constant: $\Delta \phi\st{f} = -\SI{6.0}{\milli\radian}$.
The steady-state phase shift $\Delta \phi\st{ss} = -\SI{99.5}{\milli\radian}$ for both (a,b).
(c) The time-to-first frequency shift $t\st{FP}$ calculated from the data in (a,b) as well as a series of data points with $\tau\st{s} = \SI{100}{\ns}$ and $\Delta\phi\st{f} = \SI{0}{\milli\radian}$ to $\SI{-6}{\milli\radian}$.
(d) The time-to-first frequency shift $t\st{FP}$ plotted \latin{versus} the force-induced phase shift $\Delta \phi\st{f}$.
}
\label{fig:trEFM-04}
\end{figure*}

Figure~\ref{fig:trEFM-03} shows how the modeled cantilever phase depends on $\Delta\phi\st{f}$ and the photocapacitance risetime $\tau\st{s}$.
We plot the inputs to the model in Fig.~\ref{fig:trEFM-03}(a,b) and the  cantilever phase calculated using Eq.~(\ref{eq:phi-mod-closed-form}) in Fig.~\ref{fig:trEFM-03}(c).
The orange curve shows the case where the steady-state phase shift is $\Delta \phi\st{ss} = \SI{-100}{\milli\radian}$, the force-induced phase shift is $\Delta \phi\st{f} = \SI{-3}{\milli\radian}$, and the photocapacitance risetime is $\tau\st{s} = \SI{100}{\ns}$.
The two blue curves show the effect of varying the photocapacitance risetime: $\tau\st{s} = \SI{10}{\ns}$ (light blue circles) and $\tau\st{s} = \SI{1}{\us}$ (dark blue squares).
The two green curves show the effect of varying the magnitude of the force-induced phase shift: $\Delta \phi\st{f} = \SI{-6}{\milli\radian}$ (light green triangles) and $\SI{0}{\milli\radian}$ (dark green stars).
Changing the magnitude of $\Delta \phi\st{f}$ causes large, persistent differences in the resulting phase-\latin{versus}-time data (Fig.~\ref{fig:trEFM-03}(c)).
In contrast, changing $\tau\st{s}$ by an order of magnitude causes almost no difference in the resulting phase-\latin{versus}-time data after the first few microseconds.
The small, transient differences in phase caused by changes in $\tau\st{s}$ would be even more difficult to detect after convolving with the \SI{64}{\us} FWHM low-pass lock-in amplifier or demodulation filter.
The persistent differences in modeled phase related to $\Delta \phi\st{f}$ indicate that the measured phase \latin{versus} time and calculated time-to-first-frequency-shift peak should be very sensitive to changes in $\Delta \phi\st{f}$, the magnitude of the phase shift induced by the abrupt shift in the tip-sample force.

Figure~\ref{fig:trEFM-04} illustrates how differences in phase relate to differences in the measured frequency shift $\delta f\st{meas}$ and time to first frequency shift peak $t\st{FP}$ when the sample photocapacitance risetime is faster than the inverse of the lock-in amplifier or demodulation bandwidth.
Figure~\ref{fig:trEFM-04}(a) shows $\delta f\st{meas}$ for a series of photocapacitance risetimes from \SI{1}{\ns} to \SI{1}{\us} with the magnitude of the change in cantilever tip-sample force $F_{h\nu} = - k_0 \times \SI{0.06}{\nm}$.
The time to first frequency shift peak $t\st{FP}$ becomes shorter at faster photocapacitance risetimes.
From \eqnref{eq:delta-phi-f}, we know that the magnitude of the force-induced phase shift $\Delta \phi\st{f}$ increases dramatically as $\tau\st{s}$ becomes faster than the cantilever inverse angular frequency because the exponential risetime change in force starts to contain significant content at the cantilever resonance frequency.
To illustrate the importance of this effect, we show $\delta f\st{meas}$ for a series of photocapacitance risetimes from \SI{1}{\ns} to \SI{1}{\us} with the magnitude of force-induced frequency shift held constant as $\Delta \phi\st{f} = \SI{-6}{\milli\radian}$, equivalent to $\Delta \phi\st{f}$ for the fastest photocapacitance risetimes in Fig.~\ref{fig:trEFM-04}(a).
With $\Delta \phi\st{f}$ held constant, there is very little change in $t\st{FP}$ over the range of photocapacitance risetimes.
This result is expected because the maximum bandwidth at which the cantilever amplitude and phase can be demodulated is a fraction of the cantilever's resonance frequency: $b \leq f_0/4$, for example.
For a \SI{500}{\kHz} cantilever, dynamics faster than 1 to 10 \si{\us} are significantly blurred by the demodulation filter.
For this reason, dynamics on these fast time scales are typically detected with pump-probe based techniques \cite{Hamers1990nov,Weiss1993nov,Nunes1993nov,Takeuchi2002jul,Cocker2013jul,Murawski2015oct,Schumacher2016apr}.

Fig.~\ref{fig:trEFM-04}(c) plots the time to first frequency shift peak calculated from the data in Fig.~\ref{fig:trEFM-04}(a) and (b).
We also plot in green the time to first frequency shift peak calculated by fixing the photocapacitance risetime $\tau\st{s} = \SI{100}{\ns}$ and varying the magnitude of the force-induced phase shift $\Delta \phi\st{f}$ from  \SI{0}{\milli\radian} to \SI{-6}{\milli\radian}.
Together, the blue and green curves show that $t\st{FP}$ is not a reliable measure of the sample photocapacitance risetime $\tau\st{s}$.
Figure~\ref{fig:trEFM-04}(d) plots the same $t\st{FP}$ data \latin{versus} the force-induced phase shift $\Delta\phi\st{f}$.
The three different curves from Fig.~\ref{fig:trEFM-04}(c) collapse to a single line, with $t\st{FP}$ linearly related to $\Delta \phi\st{f}$ over this range of time constants and force-induced phase shifts.
The theory and simulations indicate that for photocapacitance risetimes $\tau\st{s}$ much smaller than the inverse filter bandwidth $1/(2\pi b)$, FF-trEFM mainly detects the total magnitude of the force-induced phase shift $\Delta \phi\st{f}$.
The force-induced phase shift depends on both the magnitude of the change in force and the photocapacitance risetime.
To relate $t\st{FP}$ to a specific photocapacitance risetime, additional information must be known or assumed about the magnitude of the abrupt change in tip-sample force.

\begin{figure*}
\includegraphics{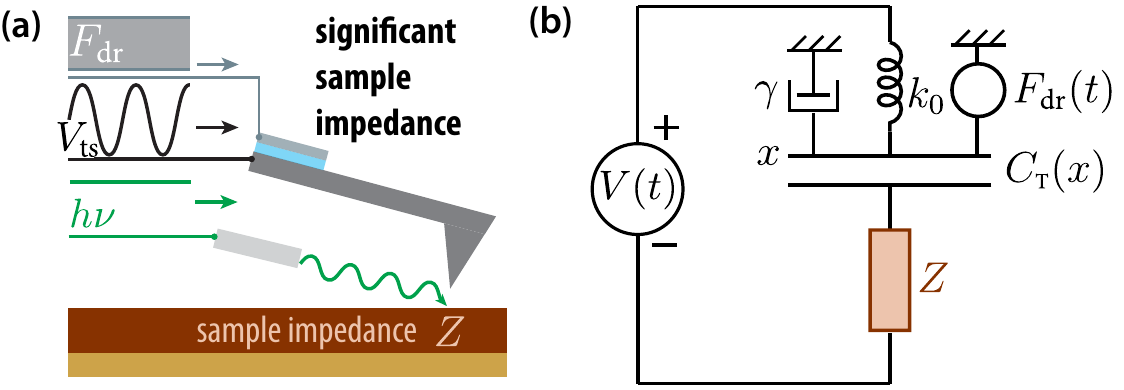}
\caption{Impedance spectroscopy model of EFM.
(a) We consider experiments at steady-state conditions with significant sample impedance $Z$.
(b) The Lagrangian model for the the experiment.
}
\label{fig:impedance-model}
\end{figure*}

To illustrate this point, we show the predicted $t\st{FP}$ by convolving the modeled phase $\phi\st{mod}$ with a low-pass lock-in amplifier filter with cutoff frequency $\omega\st{f}$. In this case, the first frequency shift peak occurs at
\begin{equation}
t\st{FP} = (\omega\st{f}-\gamma)^{-1} \log \left( \frac{\,\,\,\omega\st{f} (\tau\st{s}^{-1} - \gamma) ( \Delta \phi\st{ss} \omega\st{f} - \Delta \phi\st{f} \gamma)}{\gamma^2 (\tau\st{s}^{-1} - \omega\st{f})(\Delta \phi\st{ss} - \Delta \phi\st{f}) }  \right).
\label{eq:tFP-I}
\end{equation}
To connect this result back to experimental parameters, we expand to first order in $\tau\st{s}$ and $\Delta \phi\st{f}$ near zero, and find
\begin{equation}
t\st{FP} = (\omega\st{f}-\gamma)^{-1} \log \left ( \frac{\omega\st{f}^2}{\gamma^2} \right) + \tau\st{s} + \frac{\Delta C'_{h\nu}}{k_1 Q A_0  \omega\st{f} \Delta C''_{h\nu} (1+ \omega_1^2 \tau\st{s}^2)}.
\label{eq:tFP-II}
\end{equation}
In this limit, the time to first frequency shift peak is the sum of a constant factor related to the cantilever damping parameter (or ringdown time) and the chosen filter function, the sought-after photocapacitance risetime, and a factor that depends on the light-induced changes to the capacitance derivatives and the photocapacitance risetime. Figures~\ref{fig:trEFM-03} and \ref{fig:trEFM-04} and the previous results of Karatay and co-workers \cite{Karatay2016may} show that for small photocapacitance risetimes ($\tau\st{s}<2\pi/\omega_1$) the final term dominates and unfortunately, the measured $t\st{FP}$ depends nonlinearly on $\tau\st{s}$ with a coefficient that is sensitive to small changes in $\Delta C'_{h\nu}/\Delta C''_{h\nu}$.
In contrast, pk-EFM can detect small changes in photocapacitance risetime because the measurement indirectly senses the total cantilever phase accumulated versus time using a series of voltage and light pulses \cite{Dwyer2017jun}.
The effect of the step-like change in tip-sample force ($\Delta C'_{h\nu}$) is explicitly accounted for.
For a photocapacitance having single-exponential kinetics, the resulting phase shift for short photocapacitance risetimes is 
\begin{equation}
\Delta \phi(t\st{p}) = 
\frac{\Delta C'_{h\nu} V^2 }{2 A_0 k_1}  \frac{\omega_1}{1 + \tau\st{s}^2 \omega_1^2}
\left (
t\st{p} - \tau\st{s} + \tau\st{s} e^{-t\st{p}/\tau\st{s}}
\right ),
\end{equation}
where $t\st{p}$ is the pulse time.
By measuring the phase shift $\Delta \phi$ \latin{versus} the pulse time $t\st{p}$, the photocapacitance risetime $\tau\st{s}$ can be extracted along with $\Delta C'_{h\nu}$.

\section{Impedance spectroscopy EFM theory}
\label{sec:Impedance_spectroscopy_EFM_theory}

In this section we analyze experiments where the assumption that tip charge  responds instantaneously to changes in the tip-sample separation or voltage breaks down.
We consider steady-state measurements so the assumption that there are no abrupt changes in the tip-sample force or force-gradient is valid.
This case covers dissipation measurements \cite{Denk1991oct}, local dielectric spectroscopy (LDS) \cite{Crider2007jul,Crider2008jan}, and broadband local dielectric spectroscopy \cite{Labardi2016may, Tirmzi2017jan}.
In the literature, these experiments are normally described by assuming a time- or frequency-dependent complex capacitance, a basically phenomenological approach that fails to clearly separate the contributions of the tip and sample impedance.

When the assumption that tip change responds instantaneously breaks down, the Lagrangian equations of motion derived in Sec.~\ref{Sec:Lagrangian-introduced} are, in general, a set of coupled, nonlinear, differential algebraic equations.
As shown in the derivations of Sec.~\ref{Sec:Current-induced-dissipation} and \ref{Sec:more-realistic-EFM}, the Lagrangian equations of motion can be reduced to a set of coupled, nonlinear ordinary differential equations.
Even this simplification, however, necessitates keeping track of numerous extraneous charge variables, requires starting over if the model of the sample and wiring impedance is changed, and most importantly, retains the coupling between the evolution of the charge variables and the evolution of the tip position.
In Sec.~\ref{Sec:Current-induced-dissipation} we addressed these limitations by linearizing both the charge and displacement coordinates about some equilibrium position.
This approach is not suitable when large modulation voltages are applied, which is the case for experiments such as local dielectric spectroscopy and frequency-modulated Kelvin probe force microscopy (FM-KPFM).
Motivated by the idea that the coupling between the charge and tip position is in some sense small, we make a carefully controlled set of approximations designed to decouple the charge and tip position so that we can relate the measured observables (cantilever frequency shift and sample-induced dissipation) to the sample impedance and cantilever response function.

\begin{figure*}
\includegraphics{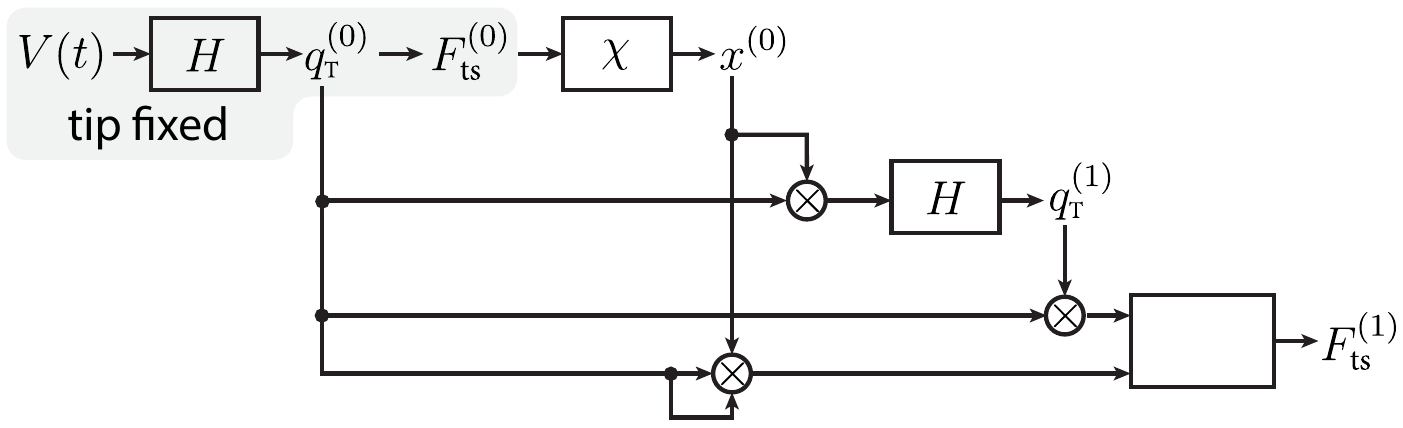}
\caption{An outline of the approximations necessary to describe EFM using impedance spectroscopy.
}
\label{fig:impedance-outline}
\end{figure*}

We start from the Lagrangian, dissipation, and generalized forces developed according to the procedure in Sec.~\ref{Sec:Lagrangian-introduced}.
We describe the sample with a general impedance $Z$ that could in principle contain any number of linear circuit elements or even impedances such as the Warburg diffusion element that cannot be expressed using only linear circuit elements \cite{Macdonald2005} (Fig.~\ref{fig:impedance-model}).
Our experimental observables are the cantilever frequency and dissipation so we focus on the equation of motion associated with the tip position $x$ (\eqnref{eq:x-eom}) and the tip-sample electro-mechanical force
\begin{equation}
F\st{ts}(q\stt{T}, x) = \frac{C'\stt{T}(x) q\stt{T}^2}{2 C\stt{T}(x)^2},
\label{eq:Fts-q-x}
\end{equation}
which depends on the cantilever tip charge and the position-dependent tip capacitance.
Equation~(\ref{eq:Fts-q-x}) is only of limited use because we need to solve a system of coupled, nonlinear differential equations to determine $q\stt{T}$ and $x$.
In Sec.~\ref{Sec:Current-induced-dissipation}, we proceeded by linearizing both the tip displacement $x$ and the tip charge $q\stt{T}$.
However, in this section, we are interested in modeling experiments that involve large amplitude, high-frequency modulations of the tip charge, so we cannot linearize the tip charge $q\stt{T}$.
Instead, we follow the series of approximations and calculations outlined in Fig.~\ref{fig:impedance-outline}.
The first assumption we make is that the tip oscillation is small so we can linearize the equations of motion in $x$. The tip-sample force then becomes
\begin{equation}
  F\st{ts} = \frac{1}{2} C' \frac{q\stt{T}^2}{C^2} + \underbrace{ \frac{1}{2} C''_q \frac{q\stt{T}^2}{C^2} x},
\label{eq:Fts}  
\end{equation}
where $C = C\stt{T}(0)$, $C' = C'\stt{T}(0)$ and 
\begin{equation}
  C''_q = C'' - 2 \frac{C'^2}{C} \propto \pdc{F\st{ts}}{x}{q\stt{T}}
  \label{eq:Czz_q} 
\end{equation}
describes the tip-sample force gradient at constant charge.
\footnote{Consider a constant applied tip-sample voltage $V$. At constant tip voltage (tip charge responds instantaneously to changes in tip position), the force gradient is $\pdc{F\st{ts}}{x}{V} = \frac{1}{2} C'' (V - \Phi)^2$. Similarly, at constant tip charge, the force gradient is $\pdc{F\st{ts}}{x}{q} = \frac{1}{2} C''_q (V - \Phi)^2$.}
The first term in \eqnref{eq:Fts} describes the force detected in amplitude-modulation EFM or KPFM experiments.
Both terms contribute to the force gradient because the charge $q\st{T}$ oscillates as the tip oscillates.
For writing experimental quantities, it is convenient to define the difference between $C''_q$ and $C''$,
\begin{equation}
\Delta C'' = 2 \frac{C'^2}{C}.
\label{eq:DeltaC}
\end{equation}

Some of the charge variables $q_{i}$ contain a term proportional to $q_{i} x$, which arises from linearizing terms involving the tip capacitance.
For example, in the simplest case where the sample impedance is purely resistive ($Z = R\stt{S}$, the case treated in Sec.~\ref{Sec:Current-induced-dissipation}), linearizing \eqnref{eq:qT-eom-Rs} gives
\begin{equation}
V(t) = \dot{q}\stt{T}R\stt{S} + \frac{q\stt{T}}{C} - \underbrace{\frac{C'}{C}\frac{q\st{T}}{C} x},
\label{eq:qT-lin}
\end{equation}
where we assume $\Phi = 0$ here and throughout this section.
The underbraced terms in Eqs.~(\ref{eq:Fts}) and (\ref{eq:qT-lin}) couple the evolution of the tip position and the tip charge.
The second assumption we make is that the coupling is small so that we can treat the underbraced terms as perturbations of order $\varepsilon$ and apply perturbation theory to dramatically simplify the system of differential equations \cite{Simmonds1997book}.
We expand the tip position $x$ in powers of $\varepsilon$:
\begin{align}
x& = x^{(0)} + \varepsilon \, x^{(1)} + \ldots ,
\end{align}
where $x^{(0)}$ is the zeroth-order approximation of the tip position and $x^{(1)}$ is the first order correction to the tip position.
Analogously, we expand the tip charge $q\stt{T}$ and any other necessary charge variables (abbreviated $q_i$) as
\begin{align}
  q\stt{T}& = q\stt{T}^{(0)} + \varepsilon \, q\stt{T}^{(1)} + \ldots ,\\
q_{i}& = q_{i}^{(0)} + \varepsilon \, q_{i}^{(1)} + \ldots .
\end{align}
By design, $q\stt{T}^{(0)}$ is independent of the tip position $x$ (see Eq.~(\ref{eq:qT-lin})).
Physically, $q\stt{T}^{(0)}$ is the tip charge assuming the tip is fixed at $x=0$.
For a given circuit and applied tip-sample voltage, we determine $q\stt{T}^{(0)}$ using ordinary circuit analysis techniques.
We are interested in experiments that probe frequency shift or dissipation at steady state, so we can neglect transients and use the transfer function between the tip voltage drop $V\st{t}$ and the applied tip-sample voltage $V$ to determine $q\stt{T}$:
\begin{equation}
\hat{H}(\omega) = \frac{\hat{V}\st{t}(\omega)}{\hat{V}\st{ts}(\omega)} = \frac{1/(j \omega C)}{Z(\omega) + 1/(j \omega C)},
\label{eq:H}
\end{equation}
where $Z$ is the sample impedance and $\hat{V}$ denotes the Fourier transform of $V$ with respect to time.
The Fourier transform of the zeroth order tip charge is
\begin{equation}
\hat{q}\stt{T}^{(0)}(\omega) = C \, \hat{H}(\omega) \hat{V}\st{ts}(\omega).
\end{equation}

Next we can determine the zeroth order cantilever position $\hat{x}^{(0)}$ which is the sum of an oscillation at frequency $\omega$ (amplitude $A_0$ determined by the driving force $F\st{dr}$) and the small oscillation induced by the zeroth order tip-sample force,
\begin{equation}
F\st{ts}^{(0)} = \frac{1}{2} \frac{C' q\stt{T}^{(0)} q\stt{T}^{(0)}}{C^2}.
\end{equation}
In the frequency domain, the additional oscillation induced by $q\stt{T}^{(0)}$ is
\begin{equation}
\hat{x}_{F\st{ts}}(\omega) = \hat{\chi}(\omega) \hat{F}\st{ts}^{(0)}(\omega),
\label{eq:x_Fts}
\end{equation}
where
\begin{equation}
  \hat{\chi}(\omega) = \frac{1}{k_0}\left ( 1 - \frac{\omega^2}{\omega_0^2} + \frac{j \omega}{Q \omega_0} \right )^{-1}
\end{equation}
is the transfer function of the oscillator
and 
\begin{equation}
\hat{F}\st{ts}^{(0)}(\omega) = \frac{1}{2} \frac{C'}{C^2} \, [\hat{q}\stt{T}^{(0)} \ast \hat{q}\stt{T}^{(0)}](\omega)
\end{equation}
is the Fourier transform of the zeroth order tip-sample force, with $\ast$ denoting convolution in the frequency domain.
We can describe EFM force measurements with just \eqnref{eq:x_Fts}.
In order to describe force-gradient measurements, we will need to compute $F\st{ts}^{(1)}$, which will re-introduce the coupling between the tip charge and tip position and cause small changes in the cantilever's amplitude, frequency and phase.

At this point we have zeroth-order approximations for the tip charge and tip position.
Next we determine the additional charge oscillation $q\stt{T}^{(1)}$ induced by the oscillating tip.
The sample impedance is unchanged and because we have assumed the tip oscillation is small, the tip capacitance during each oscillation is approximately constant. The first order tip charge is driven by the effective voltage source
\begin{equation}
  V_{x} = \frac{C'}{C^2} q\stt{T}^{(0)} x^{(0)}.
\end{equation}
The resulting first order correction to the tip charge is
\begin{equation}
  \hat{q}\stt{T}^{(1)}(\omega) = C \, \hat{H}(\omega) \hat{V}_x(\omega),
\end{equation}
where the transfer function $\hat{H}$ is given by \eqnref{eq:H} and the Fourier transform of the effective voltage source is
\begin{equation}
    \hat{V}_{x}(\omega) = \frac{C'}{C^2} \, [ \hat{q}\stt{T}^{(0)} \ast \hat{x}^{(0)}](\omega).
\end{equation}

Finally, we can determine the first-order correction to the tip-sample force, and therefore determine how the cantilever amplitude and frequency depend on sample properties and the applied modulation voltage.
Our final approximation is that this first-order correction is sufficient to approximate the cantilever frequency shift and sample-induced dissipation.
The first-order correction to the tip-sample force is
\begin{equation}
  F\st{ts}^{(1)} = \frac{C' q\stt{T}^{(0)} q\stt{T}^{(1)}}{C^2} + \frac{1}{2} C''_q \frac{q\stt{T}^{(0)} q\stt{T}^{(0)} x^{(0)}}{C^2}.
\label{eq:Fts1}
\end{equation}
It is useful to recall the limiting behavior of the tip-sample force, frequency shift, and sample-induced dissipation in the case of a purely resistive sample (Sec.~\ref{Sec:Current-induced-dissipation}).
In the limit that the sample impedance $Z(\omega_0)$ is large compared to the tip impedance $1/(j\omega_0 C)$, the tip charge remains constant throughout the oscillation cycle so that $q\stt{T}^{(1)} = 0$.
In this case, the force gradient is determined entirely by $C''_q$, which is related to the change in electric field between the tip and sample at constant charge.\footnote{Recall that for an infinite parallel plate capacitor, the electric field $E$ is independent of the distance between the plates for a fixed charge.}
The usual approximation is that the sample impedance $Z(\omega_0)$ is negligible compared to the tip impedance so that the tip charge responds instantaneously to any change in tip position.
In this case, $q\st{T}^{(1)} = C'^2 q\stt{T}^{(0)} x^{(0)} / C^2$ and the force gradient reduces to $\frac{1}{2} C'' V\st{t}(t)^2$. Both the oscillating charge induced by the oscillating tip and the effect of the oscillating tip on the electric field between tip and sample contribute to the measured force gradient.
Note as well that as long as the tip voltage is not determined from the tip position using feedback, the oscillating force caused by the second term in \eqnref{eq:Fts1} will be purely in phase with the cantilever oscillation and cause a frequency shift.
In contrast, the first term can give rise to a frequency shift or dissipation.

To determine the frequency shift and sample-induced dissipation, we first take the lock-in amplifier signal to be
\begin{equation}
F\st{LIA} = F\st{ts}^{(1)} e^{-j \omega_0 t},
\end{equation}
where we assume that the oscillation induced by the drive force is $x = A_0 \cos (\omega_0 t)$.
The real part of $\hat{F}\st{LIA}(0)$ corresponds to a force in phase with the cantilever oscillation, which causes a frequency shift 
\begin{equation}
\Delta f = -\frac{f_0}{2k_0} \frac{\Re{\hat{F}\st{LIA}(0)}}{A},
\end{equation}
while the imaginary part of $F\st{LIA}$ corresponds to a force out of phase with the cantilever oscillation, which causes a sample-induced dissipation 
\begin{equation}
\Gamma\st{s} = -\frac{1}{\omega\st{c}} \frac{\Im{\hat{F}\st{LIA}(0)}}{A} .
\end{equation}

In many experiments, we modulate the tip voltage and detect the frequency shift at some non-zero frequency.
In this case, we isolate the force component responsible for a frequency shift by taking
\begin{equation}
  F\st{in\text{-}phase} = F\st{ts} \cos(\omega_0 t) , 
\end{equation}
and then obtain the frequency shift as a function of frequency by taking the Fourier transform,
\begin{equation}
\Delta \hat{f}(\omega) = - \frac{f_0}{2k_0} \frac{\hat{F}\st{in\text{-}phase}(\omega)}{A},
\label{eq:Delta_f_omega}
\end{equation}
where $\Delta \hat{f}(\omega)$ is a complex number representing the output of both the X- and Y-channels of a lock-in amplifier set to frequency $\omega$.
Below, we apply the procedure outlined in Eqs.~(\ref{eq:H})--(\ref{eq:Delta_f_omega}) to determine the frequency shift and/or sample-induced dissipation in different experiments as a function of the sample impedance.

\subsection{Frequency and dissipation \latin{versus} voltage}

We first consider applying a constant tip-sample voltage $V\st{ts} = V + \Phi$ and driving the cantilever at its resonance frequency using a phase-locked-loop controller (PLL).\footnote{In the experiments of Ref.~\citenum{Tirmzi2017jan}, we measured the cantilever frequency and amplitude after waiting a delay time $T\st{delay} \geq 3 \tau\st{r}$, with the ringdown time $\tau\st{r} = 4 \pi f_0^{-1} Q$.
The delay allowed the cantilever time to settle to a new amplitude that reflected any dissipation caused by the tip-voltage.}
Using the procedure outlined above, we obtain the cantilever frequency shift
\begin{align}
\Delta f = -\frac{f_0}{4k_0} \Big( C''_q + \Delta C'' \Re\big(\hat{H}(\omega_0)\big) \Big ) V^2
\label{Eq:KPFM-new-major-equation}
\end{align}
and sample-induced dissipation
\begin{align}
\Gamma\st{s} = -\frac{1}{2 \omega_0} \Delta C'' \, \Im\big(\hat{H}(\omega_0) \big) \, V^2,
\end{align}
where we have assumed that the sample impedance $Z$ has a resistive component so that $\hat{H}(0) = 1$.
In the limit that the sample impedance $Z = R\stt{S}$, we recover the results derived in Sec.~\ref{Sec:Current-induced-dissipation}. 

\subsection{Local dielectric spectroscopy}
\label{sec:lds}

\begin{figure}
\includegraphics{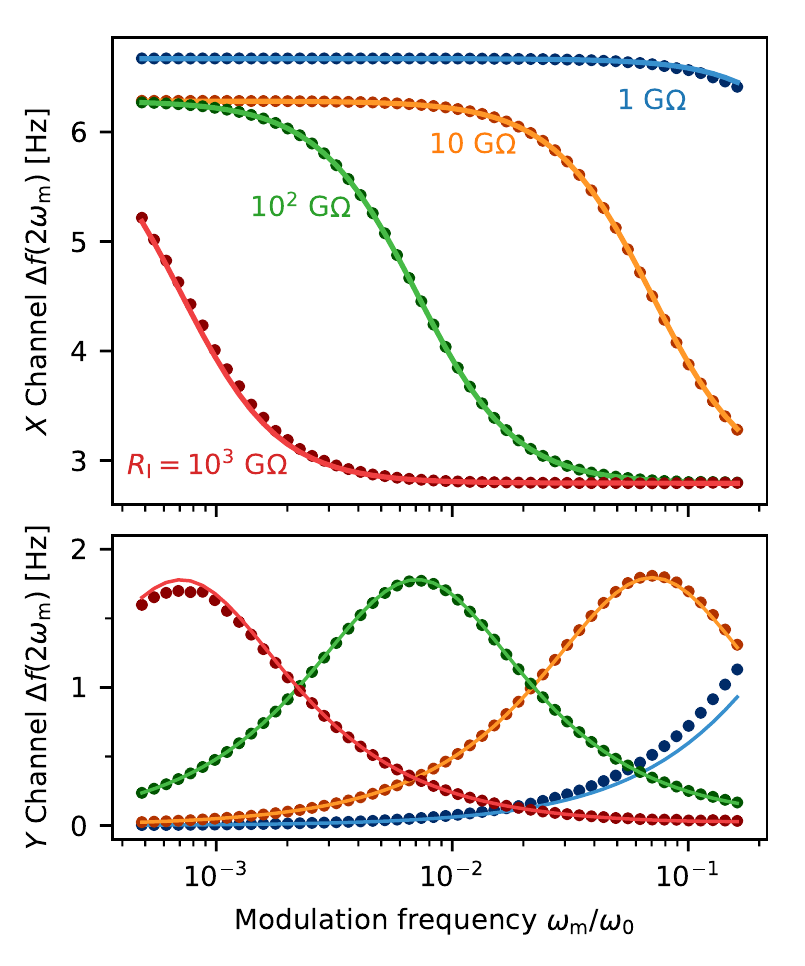}
\caption{Analytical and numerical analysis of local dielectric spectroscopy.
Comparison between frequency shift determined from numerical simulations (points) and the impedance-theory approximation (lines). Simulation parameters and details are given in Appendix~\ref{Sec:Appendix-simulation}.}
\label{fig:lds-frequency-shift}
\end{figure}

In local dielectric spectroscopy (LDS), the applied tip-sample voltage is $V\st{ts} = V\st{m} \cos(\omega\st{m} t)$ with $V\st{m}$ the modulation-voltage amplitude and $\omega\st{m}$ the modulation-voltage frequency.
The cantilever is driven at its resonance frequency using a PLL.
The cantilever frequency shift component at a frequency $2\omega\st{m}$ is monitored with a lock-in amplifier as the frequency $\omega\st{m}$ is stepped from low to high frequency, with the high frequency limit still significantly less that the cantilever resonance frequency.
The frequency shift, in this case, is found to be
\begin{equation}
\Delta \hat{f}(2\omega\st{m}) = -\frac{f_0 V\st{m}^2}{8 k_0} \Big (C''_q + \Delta C'' \bar{H}(\omega\st{m}, \omega_{0}) \Big ) \hat{H}^2(\omega\st{m}),
\label{eq:Deltaf_LDS}
\end{equation}
where $\bar{H}$ is the average response at frequencies $\omega\st{m} \pm \omega_{0}$:
\begin{equation}
\bar{H}(\omega\st{m}, \omega_{0}) = \frac{1}{2} \Big( \hat{H}(\omega\st{m} + \omega_{0}) + \hat{H}(\omega\st{m} - \omega_{0}) \Big).
\label{eq:Hbar}
\end{equation}
In LDS, $\bar{H}(\omega\st{m}, \omega_{0})$ is typically well-approximated by $\Re{\hat{H}(\omega_0)}$ because $\omega\st{m} \ll \omega_0$.
We see that the experiment mainly probes the response of sample charge at the modulation frequency $\omega\st{m}$.

To show that the first order perturbation theory approximation is good, we compare the analytic approximation of \eqnref{eq:Deltaf_LDS} to numerical simulations of the equations of motion for a sample impedance that shows dynamics over multiple timescales (Fig.~\ref{Fig:photocapacitance-EFM-circuit-2rs} and Appendix~\ref{Sec:Appendix-simulation}).
Figure~\ref{fig:lds-frequency-shift} shows the real and imaginary components of $\Delta \hat{f}(2f\st{m})$, which correspond to the outputs of the $X$- and $Y$-channels of a lock-in amplifier set to $2f\st{m}$ at various sample interfacial resistances $R\stt{I}$.
There is good agreement between the numerical simulations (points) and analytic approximation (lines) across the entire range of modulation frequencies.

\subsection{Broadband local dielectric spectroscopy}
\label{sec:blds}

\begin{figure}
\includegraphics{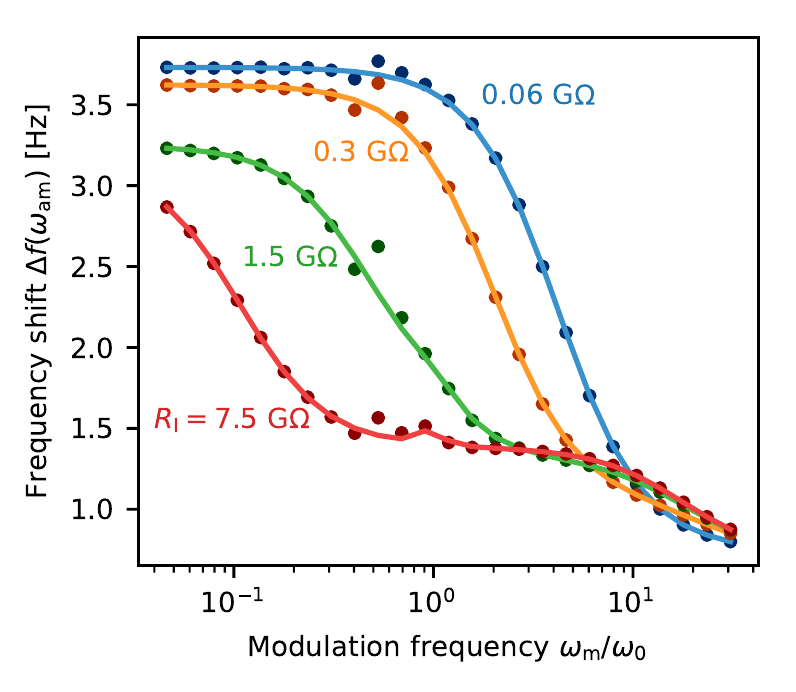}
\caption{Analytical and numerical analysis of broadband local dielectric spectroscopy. Comparison between frequency shift determined from numerical simulations (points) and the impedance-theory approximation (lines).
Simulation parameters and details are given in Appendix~\ref{Sec:Appendix-simulation}.}
\label{fig:blds-frequency-shift}
\end{figure}

While local dielectric spectroscopy probes sample charge at frequencies $\omega\st{m} \ll \omega_0$, broadband local dielectric spectroscopy (BLDS) probes the response of sample charge at higher frequencies by exploiting the nonlinear relationship between applied voltage and frequency shift to mix a high frequency signal to a convenient intermediate frequency.
In BLDS, the tip-sample voltage is
\begin{equation}
V(t)= V\st{m} \left (\frac{1}{2} + \frac{1}{2} \cos(\omega\st{am} t) \right ) \cos(\omega\st{m} t),
\label{eq:V-blds}
\end{equation}
with $V\st{m}$ the modulation-voltage amplitude, $\omega\st{m}$ the modulation-voltage frequency, and $\omega\st{am}$ the amplitude-modulation frequency.
The amplitude-modulation frequency $\omega\st{am}$ is a convenient intermediate frequency; it must be within the PLL frequency detection bandwidth ($\omega\st{am}/(2\pi) < \SI{1}{\kHz}$, typically).
The cantilever is driven at its resonance frequency using a PLL.
The cantilever frequency shift component at a frequency $\omega\st{am}$ is monitored with a lock-in amplifier as the modulation frequency $\omega\st{m}$  is stepped from low to high frequency. The frequency shift is
\begin{equation}
\Delta \hat{f}(\omega\st{am}; \omega\st{m}) = -\frac{f_0 V\st{m}^2}{16 k_0} \Big [
C''_q + \Delta C'' \Re \big( \bar{H} (\omega\st{m}, \omega_{0}) \big ) 
\Big ]
\abs{\hat{H}(\omega\st{m})}^2,
\label{eq:Deltaf-BLDS}
\end{equation}
where $\bar{H}$ is the average response at frequencies $\omega\st{m} \pm \omega_{0}$ (\eqnref{eq:Hbar}).
In contrast to LDS, which retains information about both the real and imaginary components of the sample response at the modulation frequency, in BLDS the frequency mixing necessary to measure the response of sample charge at high frequencies results in the loss of phase information.

Just as for LDS, we compare the analytic approximation for the BLDS frequency shift in \eqnref{eq:Deltaf-BLDS} to numerical simulations of the equations of motion (Fig.~\ref{Fig:photocapacitance-EFM-circuit-2rs} and Appendix~\ref{Sec:Appendix-simulation}).
Figure~\ref{fig:blds-frequency-shift} shows $\Delta \hat{f}(f\st{am})$ versus the applied modulation frequency at various sample interfacial resistances $R\stt{I}$.
There is good agreement between the numerical simulations (points) and analytic approximation (lines) across the entire range of modulation frequencies.
Overall, the procedure outlined in this section provides a way to analyze \emph{any} steady-state force or force-gradient measurement for an arbitrary sample impedance.

\subsection{Parallel resistance and capacitance sample impedance}
Just as in impedance spectroscopy, a specific model of the sample impedance is needed to extract relevant information about the sample from these experiments.
Here we describe a model that was useful in interpreting the experimental results in Ref.~\citenum{Tirmzi2017jan}.
We model the sample using a parallel resistance $R\stt{S}$ and capacitance $C\stt{S}$ so that the sample impedance $Z = (R\stt{S}^{-1} + j \omega C\stt{S})^{-1}$.
In this case, the transfer function $\hat{H}$ is
\begin{equation}
\hat{H}(\omega) = \frac{R\stt{S}C\stt{S} \omega - j}{R\stt{S}(C\stt{S} + C\stt{T}) \omega - j}.
\label{eq:H-RsCs}
\end{equation}
The circuit is a lag compensator with time constant $\tau$ and gain parameter $g$ given by, respectively,
\begin{equation}
\tau = R\stt{S} (C\stt{S} + C\stt{T})  \hspace{0.5em} \text{and} \hspace{0.5em} g = (C\stt{S} + C\stt{T}) / C\stt{S}.
\label{eq:tau-g}
\end{equation}

This model provides an intuitive way to interpret the representative BLDS data of Fig.~\ref{fig:BLDS-intro}.
As the light intensity is increased (from bottom dark points to top light points), the decrease in $\Delta f$ occurs at higher modulation frequencies, indicating that the time constant $\tau$ decreases as the light intensity increases.
According to \eqnref{eq:tau-g}, the decrease in $\tau$ could be caused by a decrease in sample resistance $R\stt{S}$ or a decrease in sample capacitance $C\stt{S}$.
We can distinguish between these two possibilities by examining the limiting behavior at high frequencies.
According to \eqnref{eq:H-RsCs}, at high frequencies the transfer function $H$ approaches $C\stt{S}/(C\stt{S} + C\stt{T})$.
In the dark, $\Delta f$ approaches zero at high frequencies, which indicates $C\stt{S} \ll C\stt{T}$.
As a result, the time constant is relatively insensitive to changes in the sample capacitance.
Consequently, $\tau \approx R\stt{S} C\stt{T}$.
We can therefore ascribe the decrease in $\tau$ with increasing light intensity to a decrease in sample resistance $R\stt{S}$.
This conclusion is robust to the sample impedance model used because even for a more complicated sample impedance model, any resistances behave as an open-circuit at high frequencies, and in the high frequency limit, only the  capacitance across the sample would be important.
Moreover, the careful analysis of the BLDS sheds light on the frequency shift and dissipation \emph{versus} light intensity data in Fig.~\ref{fig:Tirmzi2017jan-expt}(c) and (d).
At the point of maximum dissipation, $R\stt{S} C\stt{T} = \omega_0^{-1}$.
We can use dissipation as a measure of the local sample resistance (or conductivity) \cite{Denk1991oct}.
The clear separation of the tip capacitance $C\stt{T}$ from the sample impedance $Z$ is a major advantage of the method presented here.
In Ref.~\citenum{Tirmzi2017jan}, the model helped us relate changes in sample time constant and dissipation to light-induced changes in the sample conductivity.

 \section{Conclusions}
\label{Sec:Discussion}

The usual description of the EFM experiment (Eqs.~(\ref{eq:F-SKPM-const-V}) and (\ref{eq:Df-SKPM-const-V})) implicitly assumes that tip charge redistributes instantaneously as the tip oscillates and that the tip-sample force and force-gradient do not change abruptly.
In the Introduction we summarized a broad range of experiments where these assumptions are violated.
To lift these assumptions we turned to Lagrangian mechanics to describe the coupled motion of the tip charge, tip coordinate, and sample charge.
The resulting coupled differential equations are exact but nonlinear and insoluble; we linearize these equations to obtain an approximate closed-form solution.
This linearization is a good approximation in the limit of small-amplitude charge and position oscillations about equilibrium.
Moving beyond this approximation would bring in nonlinear oscillator physics such as an amplitude-dependent frequency and bistability.
In contrast to high-resolution AFM experiments, these nonlinear effects have not, to our knowledge, been significant in most high-sensitivity EFM experiments carried out to date.
The Lagrangian-mechanics approach outlined here is nevertheless an excellent starting point for treating nonlinear effects in electrical scanning-probe experiments.
Nonlinear terms would have to be measured experimentally, however, and the associated nonlinear equations of motion solved numerically.
Within the small-amplitude approximation, we have developed an analytical framework (Sec.~\ref{Sec:Lagrangian-introduced}) of closed-form equations for interpreting a broad range of EFM measurements where the usually employed but often inapplicable adiabatic-charge-redistribution and abrupt-$\Delta k$-change assumptions are violated.

Sections~\ref{Sec:Magnus} to \ref{Sec:FF-trEFM} show how our framework can be used to quantitatively analyze photocapacitance measurements that involve abrupt changes in the tip-sample force and force gradient when the light intensity or applied voltage is abruptly changed.
We derived how an abrupt change to the tip-sample force induces an abrupt change in the cantilever's amplitude and phase, and used this result to obtain a new analytical expression for the $t\st{FP}$ observable in the FF-trEFM experiment.
Taken together with our prior analysis of the pk-EFM experiment \cite{Dwyer2017jun}, we see that the results of Sections~\ref{Sec:Magnus} to \ref{Sec:FF-trEFM} give us a framework for fully evaluating the sub-cycle time resolution of ultrafast electrical scanning-probe experiments.
While we focused on the photocapacitive effects that were most important for understanding the relevant experimental results \cite{Giridharagopal2012jan,Karatay2016may,Dwyer2017jun}, our analysis also applies to situations where the dominant factor is light- or voltage-induced changes in the sample's surface potential \cite{Takihara2008jul,Shao2014oct,Murawski2015oct,Murawski2015dec,Schumacher2016apr}. 

In Sec.~\ref{sec:Impedance_spectroscopy_EFM_theory}, we introduced a procedure to relate frequency shift and/or sample-induced dissipation during steady-state EFM experiments to an arbitrary sample impedance $Z$.
This procedure helped us analyze frequency shift, dissipation, local dielectric spectroscopy, and broadband local dielectric spectroscopy measurements using a common framework (Fig.~\ref{fig:overview}(d--f)).
The primary finding of Sec.~\ref{sec:Impedance_spectroscopy_EFM_theory} is that \eqnref{eq:Df-SKPM-const-V}, ubiquitously employed to describe the FM-KPFM experiment, should be replaced by \eqnref{Eq:KPFM-new-major-equation} when interrogating any sample having finite resistance and capacitance.
While we focused in Sec.~\ref{sec:Impedance_spectroscopy_EFM_theory} on analyzing light-induced changes to the sample impedance, the model could also accommodate light-induced changes to the surface potential $\Phi$ or describe how the sample impedance would impact novel Kelvin probe force microscopy measurements such as heterodyne KPFM \cite{Sugawara2012may,Garrett2016jun}, dissipative KPFM \cite{Miyahara2015nov,Miyahara2017apr}, or open-loop KPFM \cite{Takeuchi2007aug,Collins2013nov} which seek to combine the spatial resolution of force-gradient measurements with the temporal resolution of force measurements.
Our approach reveals how the signal in these experiments changes when the sample impedance becomes significant.

The general approach outlined in Sec.~\ref{Sec:Lagrangian-introduced} and Sec.~\ref{sec:Impedance_spectroscopy_EFM_theory} provide another possible, rigorous route to describe the tip-sample interaction and cantilever parameters in piezoresponse and electrochemical strain microscopy.
In this case, an electromechanical model of the sample, with a sample displacement variable, would be necessary. With such a model, the Lagrangian formalism could be used to generate the coupled equations of motion and the tip-sample force, frequency shift, and friction could be derived.

The impedance theory description of EFM also provides an interesting perspective on the photocapacitance experiments discussed in the previous section.
In the context of that theory, an apparent increase in capacitance could be caused by an increase in sample capacitance and/or a decrease in sample resistance.
Just as the combination of dissipation and broadband local dielectric spectroscopy was informative for the perovskite materials of Ref.~\citenum{Tirmzi2017jan}, performing photocapacitance measurement in tandem with broadband local dielectric spectroscopy on the organic bulk heterojunction films of Refs.~\citenum{Giridharagopal2012jan} and \citenum{Dwyer2017jun} could help resolve the origin of the photocapacitance signal.

\begin{acknowledgments}
The numerical simulations and analysis are available online \cite{Dwyer2018jul_data}.
The authors acknowledge Ali Tirmzi and Tobias Hanrath for the data in Fig.~\ref{fig:Tirmzi2017jan-expt} and Fig.~\ref{fig:BLDS-intro}, and Ali Tirmzi and Roger Loring for fruitful discussions.
R.P.D.\ and J.A.M.\ acknowledge the financial support of Cornell University and the U.S.\ National Science Foundation (grants NSF-DMR 1309540 and NSF-DMR 1709879).
L.E.H.\ acknowledges financial support provided by the U.S.\ Military Academy Department of Physics and Nuclear Engineering and the U.S.\ Military Academy Academic Enrichment Program.
The views expressed herein are those of the authors and do not reflect the position of the Department of the Army or the Department of Defense.
\end{acknowledgments}
 \appendix
\section{Impedance spectroscopy simulations}
\label{Sec:Appendix-simulation}

\begin{figure}
\includegraphics{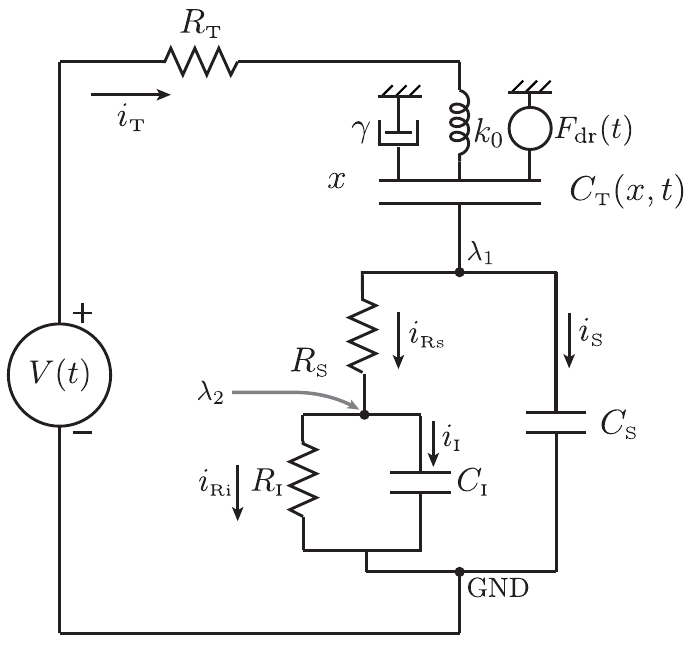}
\caption[Perovskite-like $RC$ circuit]{A circuit illustrating the types of dynamics observed in perovskite and organic-semiconductor solar cells.
}
\label{Fig:photocapacitance-EFM-circuit-2rs}
\end{figure}

This appendix lays out how the simulations shown in Sec.~\ref{sec:Impedance_spectroscopy_EFM_theory} were performed.
First we applied the procedure of Sec.~\ref{Sec:Lagrangian-introduced} to the circuit shown in Fig.~\ref{Fig:photocapacitance-EFM-circuit-2rs}.
This procedure generates eight equations: one for the tip position $x$; five for the charge variables $q\stt{T}, q\stt{S}, q\stt{I}, q\sRi,$ and $q\sRs$; and two for the Lagrangian multipliers $\lambda_1$ and $\lambda_2$.
In the limit that the tip resistance $R\st{T}$ approaches zero, there are two differential equations and five algebraic equations for the charge variables and Lagrangian multipliers. 
We reduce the dimensionality of the system by solving for $q\stt{T}, q\sRi, q\sRs, \lambda_1, \lambda_2$ in terms of the remaining variables $q\stt{S}, q\stt{I}$.
The eight differential and algebraic equations are reduced to three differential equations:
\begin{align}
m\ddot{x}& = -m \omega_0 x - 2\gamma m \dot{x} + \frac{C'\stt{T}(x)q\stt{T}^{2}}{2C^2\stt{T}(x)} + F\st{dr}(t), \\
R\stt{I} \dot{q}\sRi& = -\frac{q\sRi}{C\stt{I}} + \frac{q\sRs}{C\stt{I}}, \,\, \text{ and} \\
R\stt{S} \dot{q}\sRs& = -\left (\frac{1}{C\stt{S} + C\stt{T}(x)} + \frac{1}{C\stt{I}} \right) q\sRs + \frac{q\sRi}{C\stt{I}} + \frac{C\stt{T}(x)}{C\stt{S} + C\stt{T}(x)} V(t),
\intertext{with the tip charge given by}
q\stt{T}& = \frac{C\stt{T}(x)}{C\stt{S} + C\stt{T}(x)}q\sRs + \frac{C\stt{S} C\stt{T}(x)}{C\stt{S} + C\stt{T}(x)} V(t).
\end{align}
So far we have re-written our equations of motion in a form that will be easier to simulate numerically but have not introduced any approximations.
The equations of motion above were linearized in $x$ about $x=0$.
The resulting equations of motion, shown below, were used in the simulations:
\begin{widetext}
\begin{align}
    \dot{x}& = p/m, \\  
    \dot{p}& = -m \omega_0^2 x - 2 \gamma p + \underbrace{\frac{C' q\stt{T}^2}{2 m C^2} + \varepsilon  \frac{C''_q q\stt{T}^2 x}{2 m C^2}}_{F\st{ts}/m} + \frac{F\st{dr}(t)}{m}, \\
    \dot{q}\sRi& = -\frac{q\sRi}{C\stt{I}R\stt{I}} + \frac{q\sRs}{C\stt{I} R\stt{I}}, \\
    \dot{q}\sRs& = - \left (\frac{1}{C\stt{S} + C} - \varepsilon \frac{C' x}{(C\stt{S} + C)^2} +  \frac{1}{C\stt{I}} \right) \frac{q\sRs}{R\stt{S}} + \frac{q\sRi}{R\stt{S} C\stt{I}} + \left (\frac{C}{C\stt{S} + C} + \varepsilon \frac{C\stt{S} C' x}{(C\stt{S} + C)^2} \right)  \frac{V(t)}{R\stt{S}}.
 \end{align}
 \end{widetext}
 For both the LDS simulations of Fig.~\ref{fig:lds-frequency-shift} and the BLDS simulations of Fig.~\ref{fig:blds-frequency-shift}, the cantilever mechanical parameters were the spring constant $k_0 = \SI{3.5}{\micro\N\per\um}$, quality factor $Q = \num{26000}$, and angular resonance frequency $\omega_0 = 2 \pi \times \SI{0.065}{\mega\Hz}$, so that the cantilever mass was $m = \SI{21.0}{\ng}$, and the linear damping parameter was $\gamma = \SI{7.85e-6}{\per\us}$.
 The drive force was $F\st{dr}(t) = 0$ and the initial cantilever amplitude was $A_0 = \SI{0.05}{\um}$.
 The tip-sample capacitance parameters were $C\stt{T}(x=0) = C = \SI{1e-3}{\pF}$, $C' = \SI{-1.80e-4}{\pF\per\um}$, $C'' = \SI{1.3e-4}{\pF\per\um\squared}$ so that $C''_q = \SI{6.5e-5}{\pF\per\um\squared}$ (\eqnref{eq:Czz_q}).
 As defined in Fig.~\ref{Fig:photocapacitance-EFM-circuit-2rs}, the sample impedance parameters were $R\stt{S} = \SI{200}{\mega\ohm}$, $C\stt{S} = \SI{1e-3}{\pF}$, and $C\stt{I} = \SI{1e-3}{\pF}$.
 The value of the resistance $R\stt{I}$ is given next to each trace in Figs.~\ref{fig:lds-frequency-shift} and \ref{fig:blds-frequency-shift}.
 The given units were those used in the simulation.
 For LDS, the applied tip-sample voltage was $V(t) = V\st{m} \sin(\omega\st{m} t)$ with the modulation voltage $V\st{m} = \SI{5}{\V}$.
 For BLDS, the applied tip-sample voltage was $V(t) = V\st{m} (\frac{1}{2} + \frac{1}{2} \cos{\omega\st{am} t}) \sin(\omega\st{m} t)$, with $V\st{m} = \SI{5}{\V}$ and the amplitude-modulation frequency $\omega\st{am} = 2 \pi \times \SI{160e-6}{\MHz}$ (Eq.~(\eqnref{eq:V-blds})).

 The simulations were performed in Python using the odeint method in Scipy \cite{Jones2001}, which calls the LSODE solver.
 Each LDS experiment was simulated for \SI{40000}{\us} and each BLDS experiment was simulated for \SI{20000}{\us}.
 The initial cantilever state was $x =A_0, p = 0$.
 Simulation transients were avoided by defining the initial charge variables $q\sRi$ and $q\sRs$ using the appropriate response function:
 \begin{align}
 q\sRs =  V \frac{C + C C\stt{I} R\stt{I} s}{1 + (R\stt{I} + R\stt{S}) C\st{TS} s + 
 R\stt{I}  C\stt{I} s (1 + R\stt{S} C\st{TS} s)}
 \intertext{and}
 q\sRi = V \frac{C}{1 +  (R\stt{I} + R\stt{S})C\st{TS} s + 
  R\stt{I} C\stt{I} s (1 +  R\stt{S} C\st{TS} s)},
 \end{align}
 where $s = j \omega$, $C\st{TS} = C+ C\stt{S}$, and the charges at $t=0$ are determined by setting $V = V\st{m} \exp (j \omega t - j \pi/2)$ and evaluating the real part of $q\sRs$ and $q\sRi$ at $t=0$ for $\omega = \omega\st{m}$.
While LSODE controls the integration method, order, and step size, inspection of the full output of the solver showed that a 5th order backward differentiation formula (BDF) Gear method was typically used with time steps of approximately \SI{0.2}{\us}.

\bibliographystyle{grants-ARO}
\bibliography{Dwyer201902__Lagrangian_final}

\end{document}